## Technical note

# Hydrological time series forecasting using simple combinations: Big data testing and investigations on one-year ahead river flow predictability

Georgia Papacharalampous[1,*], and Hristos Tyralis[2]

[1] Department of Water Resources and Environmental Engineering, School of Civil Engineering, National Technical University of Athens, Heroon Polytechneiou 5, 157 80 Zographou, Greece; papacharalampous.georgia@gmail.com, gpapacharalampous@hydro.ntua.gr; https://orcid.org/0000-0001-5446-954X

[2] Air Force Support Command, Hellenic Air Force, Elefsina Air Base, 192 00 Elefsina, Greece; Department of Water Resources and Environmental Engineering, School of Civil Engineering, National Technical University of Athens, Heroon Polytechneiou 5, 157 80 Zographou, Greece; montchrister@gmail.com, hristos@itia.ntua.gr; https://orcid.org/0000-0002-8932-4997

* Correspondence: papacharalampous.georgia@gmail.com, tel: +30 69474 98589

**Abstract:** Delivering useful hydrological forecasts is critical for urban and agricultural water management, hydropower generation, flood protection and management, drought mitigation and alleviation, and river basin planning and management, among others. In this work, we present and appraise a new simple and flexible methodology for hydrological time series forecasting. This methodology relies on (a) at least two individual forecasting methods and (b) the median combiner of forecasts. The appraisal is made by using a big dataset consisted of 90-year-long mean annual river flow time series from approximately 600 stations. Covering large parts of North America and Europe, these stations represent various climate and catchment characteristics, and thus can collectively support benchmarking. Five individual forecasting methods and 26 variants of the introduced methodology are applied to each time series. The application is made in one-step ahead forecasting mode. The individual methods are the last-observation benchmark, simple exponential smoothing, complex exponential smoothing, automatic autoregressive fractionally integrated moving average (ARFIMA) and Facebook's Prophet, while the 26 variants are defined by all the possible combinations (per two, three, four or five) of the five afore-mentioned methods. The new methodology is identified as well-performing in the long run, especially when more than two individual forecasting methods are combined within its framework. Moreover, the possibility of case-informed integrations of diverse hydrological forecasting methods within systematic frameworks is algorithmically investigated and discussed. The related investigations encompass linear regression analyses, which aim at finding interpretable relationships between the values of a representative forecasting performance metric and the values of selected river flow statistics. We find only loose (but not negligible) relationships between the formed variable sets. These relationships could be exploited for improving (to some extent) future forecasting applications. The results of our big data forecasting experiment are lastly exploited for characterizing –in relative terms– one-year ahead river flow predictability. We believe that our findings have both practical and theoretical implications.

## 1. Introduction

1.1 Statistical analyses and predictive modelling of hydrological processes

Understanding and modelling hydrological phenomena and their changes in time are placed among the most important considerations and challenges for both hydrological science and engineering hydrology (see Montanari et al. 2013; Blöschl et al. 2019). Offering solid grounds for progressing our understanding of changes in general, hydrological changes have been studied and explored since the early beginnings of science (Koutsoyiannis 2013). Moreover and along with this theoretical viewpoint and research orientation, interpreting and characterizing changing behaviours and patterns in hydrological process regimes through thorough hydrological analyses and model-fitting investigations can naturally and traditionally guide many of our modelling choices and methodological assumptions, thereby contributing to uncertainty reduction within modelling frameworks in hydrology. Here the interest is in predictive modelling of hydrological processes.

Based on Shmueli (2010), one should recognise and value both the practical and scientific aspects of predictive modelling of hydrological processes. The practical value is easy-to-perceive. Useful hydrological predictions (e.g., river flow forecasts) are required both at large and fine time scales for optimizing urban and agricultural water management, hydropower generation, flood protection and management, drought mitigation and alleviation, and river basin planning and management, among others (see e.g., Pulwarty and Redmond 1996; Georgakakos et al. 1998; de Roo et al. 2003; Feldman and Ingram 2009; Buizer et al. 2016). Optimal management practices in engineering hydrological contexts can, in their turn, greatly benefit the environment and the society, ensuring both environmental sustainability and economic growth. On the contrary, sub-optimal choices can have severe (and sometimes irreversible) environmental and sociological impact (see e.g., the relevant discussions in Koppel 1987; Barrow 1998; Bazilian et al. 2011; Wichelns 2017; Liu et al. 2018; Talukder and Hipel 2020).

Even when predictive modelling is data-driven, scientific value is identified –along with practical value– e.g., in (i) understanding predictive performance patterns by investigating their possible relationships with process behaviours and patterns, and (ii) acquiring knowledge on process predictability across various locations of the globe. In spite of the interest captured in point (i) above, predictive performance assessments (see e.g., Carlson et al. 1970; Jayawardena and Fernando 1998; Sivakumar et al. 2002; Chau et al. 2005; Chau 2007; Koutsoyiannis et al. 2008; Sivakumar and Berndtsson 2010; Abbas and Xuan 2019; Aguilera et al. 2019; Khatami et al. 2019; Papacharalampous et al. 2019b; Tyralis et al. 2019a) and time series analyses (see e.g., Yevjevich 1987; Montanari et al. 1997, 1999, 2000; Koutsoyiannis 2002; Villarini et al. 2011; Montanari 2012; Toth 2013; Markonis et al. 2018; Tyralis et al. 2018; Mackay et al. 2019; Steirou et al. 2019; Nerantzaki and Papalexiou 2019; Tyralis et al. 2019c; Markonis and Strnad 2020) are usually carried out independently in the literature. On the contrary, investigations related to point (ii) above are often in the literature (see e.g., Mauer and Lettenmaier 2003; Wilby et al. 2004; Papacharalampous et al. 2018). The reader is also referred to Koutsoyiannis (2010, 2013), for interesting theoretical discussions on hydrological process predictability.

Predictive performance is, in principle, appraised by computing the values of various performance metrics (see e.g., the metrics investigated in Krause et al. 2005) on data that have not been exploited for model training. These values can further be used for characterizing the examined processes in terms of predictability. Investigations related to point (i) above could be made by exploiting the metric values (computed within prediction experiments) together with selected hydrological statistics within hydrological time series analysis contexts. Hydrological statistics (else referred to as "hydrological signatures" in catchment hydrology; Tyralis et al. 2019c) are of major interest to hydrological scientists (see e.g., Villarini et al. 2011; Toth 2013; Markonis et al. 2018; Tyralis et al. 2018; Mackay et al. 2019; Tyralis et al. 2019c). A common theme characterizing studies devoted to hydrological statistics is the investigation of relationships between the examined statistics, as well as between these statistics, and catchment location and attributes.

1.2 Building on past experience in (hydrological) time series forecasting

We forecast mean annual river flow time series by exclusively using information about their past (i.e., endogenous predictor variables). The statistical and data-driven hydrological literature presents numerous case studies issuing and assessing this type of time series forecasts (see e.g., Sivakumar et al. 2002; Koutsoyiannis et al. 2008; Aguilera et al. 2019). The latter are accurate enough when delivered at large time scales (i.e., the annual, seasonal and monthly ones), while at fine time scales (i.e., the daily and sub-daily ones) exogenous predictor variables (e.g., observed or forecasted values of various hydrometeorological variables) can be very informative and, thus, their consideration can result in large improvements in forecasting performance (see e.g., the incorporation of exogenous predictors in the frameworks by Quilty et al. 2016; Quilty and Adamowski 2018). Such works have been categorized according to their primary focus, the hydrometeorological process examined, the data level and the forecast horizon by Papacharalampous et al. (2019a). Here we perform one-step ahead forecasting.

Predictive modelling can be reliably advanced under the "big data" approach (term used here as opposed to the conventional case study approach, which exploits a much smaller amount of real-world data). The big data approach is especially meaningful when predictive modelling is data-driven (for relevant discussions, see Boulesteix et al. 2018); yet, big data time series forecasting is rarely performed in the geoscientific literature (see e.g., Papacharalampous et al. 2018; Papacharalampous et al. 2019a) and even beyond geoscience (see e.g., Carta et al. 2019; Petropoulos and Svetunkov 2020). Here we follow this reliable approach by exploiting a river flow dataset compiling 90-year long information from approximately 600 stations. The exploited stations are mostly located in two continental-scale regions, specifically North America and Europe. The dataset represents various climate and catchment characteristics and is, thus, ideal for benchmarking purposes.

Time series forecasting using individual models can offer a certain degree of accuracy that could be improved through forecast combinations (see e.g., the classical work by Bates and Granger 1969, and the reviews by Granger 1989; Wallis 2011; Sagi and Rokach 2018). Here we are explicitly interested in simple combinations. Despite their simplicity, these combinations are known to outperform in many forecasting problems more complex combination methodologies (see e.g., the related discussions in De Gooijer and Hyndman 2006; Smith and Wallis 2009; Lichtendahl et al. 2013; Graefe et al. 2014; Hsiao and Wan 2014; Winkler 2015; Claeskens et al. 2016). In fact, building a sophisticated combiner that beats a simple combiner is that puzzling, that is often referred to in the literature as the "forecast combination puzzle" (see e.g., Smith and Wallis 2009; Claeskens et al. 2016). Forecast combination methodologies are a new topic in hydrological time series forecasting.

Despite their varying characteristics and statistical-modelling-culture traits (for the latter, see Breiman 2001), stochastic models (see e.g., Wei 2006; Hyndman et al. 2008) and machine learning algorithms (see e.g., Hastie et al. 2009; Alpaydin 2010; James et al. 2013) have been found to be mostly equally competitive in hydrological time series forecasting at large time scales (see the big data forecasting investigations of Papacharalampous et al. 2019a and its large-scale companion works, as defined therein). Therefore, for our simple combinations we exploit models from both these families. From the stochastic family, we implement an autoregressive fractionally integrated moving average (ARFIMA) forecasting model by Hyndman and Khandakar (2008), simple exponential smoothing (SES) by Brown (1959; see also Hyndman et al. 2008), and complex exponential smoothing (CES) by Svetunkov and Kourentzes (2016).

The ARFIMA process (Granger and Joyeux 1980; Hosking 1981) is an extension of the autoregressive integrated moving average (ARIMA) process by Box and Jenkins (1970). Being an analogous of the fractional Gaussian noise process (Kolmogorov 1940; Hurst 1951; Mandelbrot and Van Ness 1968; see also the popularizing work by Koutsoyiannis 2002, the dataset explorations in Papacharalampous et al. 2018, 2019a, and the big data predictive modelling application in Tyralis et al. 2018), ARFIMA is appropriate for modelling long-range dependence. AR(F)IMA models have been widely applied in stochastic hydrology for several purposes including (but not limited to) hydrological forecasting (see e.g., Carlson et al. 1970; Yevjevich 1987; Montanari et al. 1996, 1997, 2000; Papacharalampous et al. 2018, 2019a; see also the overview by Sivakumar 2016, Chapter 3). On the contrary, SES has only been exploited in a limited number of hydrological forecasting studies (e.g., in Papacharalampous et al. 2018, 2019a) and CES has not been applied so far for hydrological forecasting.

As a new representative of the machine learning family of time series forecasting models, we here apply Facebook's Prophet by Taylor and Letham (2018). Successful (hydrological) applications of this model can be found in Papacharalampous et al. (2018), Aguilera et al. (2019), Belikov et al. (2019), Fernández-Ayuso et al. (2019), and Yan et al. (2019). In spite of the emphasis that we here place on stochastic models, Prophet also holds an important position within our big data investigations, since machine learning algorithms are increasingly popular in applied and data-driven hydrology (see e.g., the works by Sahoo et al. 2019a,b; Srikanth et al. 2019; see also the reviews by Solomatine and Ostfeld 2008; Maier et al. 2010; Raghavendra and Deka 2014; Tyralis et al. 2019b).

## 1.3   Aims and novelties of the present work

Our aims are to: (1) introduce a simple combination methodology for hydrological time series forecasting (relying on at least two individual time series forecasting methods and the median combiner of forecasts); (2) extensively test the performance of the new methodology by exploiting a big dataset and by complying with the principles of forecasting (see e.g., Armstrong 2001; Hyndman and Athanasopoulos 2018); (3) assess the performance of CES in hydrological time series forecasting contexts; (4) compare the performance of various time series forecasting methods –emphasizing on stochastic ones– in one-step ahead annual river flow forecasting; (5) investigate the existence of possible relationships between time series forecasting performance and selected river flow statistics; and (6) characterize –in interpretable terms– one-year ahead river flow predictability within two continental-scale case studies. These aims have both practical and theoretical orientation. Each of them is also directly related to one of the six novel points characterizing the present work. These novel points are the following:

o   We introduce one of the first forecast combination methodologies for hydrological time series forecasting. This new family of hydrological time series forecasting

- o  We perform one of the most systematic empirical assessments ever made on simple combinations even beyond hydrology (and geoscience).
- o  CES is here implemented for the first time in the hydrological time series forecasting literature.
- o  Several forecasting methods are here compared for the first time in big data one-step ahead annual river flow forecasting, while placing emphasis on stochastic (rather than on soft-computing) methods is rare in the most recent hydrological time series forecasting literature.
- o  Investigating the existence of possible links between forecasting performance and statistical characteristics of the time series is a new practice in hydrology.
- o  One-step ahead annual river flow predictability is here investigated for the first time by exploiting a big dataset according to the forecasting principles and practice.

## 2. Data and methods

In this section, we present the experimental data and methods of the study. Statistical software information is independently provided in Appendix A.

### 2.1 River flow time series

We exploit information from the Global Streamflow Indices and Metadata archive, made available by Do et al. (2018b) and Gudmundsson et al. (2018b). Documentations of this archive can be found in Do et al. (2018a) and Gudmundsson et al. (2018a). From the entire archive, we first retrieve all mean annual river flow time series that simultaneously satisfy the following two conditions: (i) they are at least 90-year-long, and (ii) they pass the four homogeneity tests described in Gudmundsson et al. (2018a, Section 4.1.1). We are interested in homogeneity, since abrupt changes (indicated by non-homogenous behaviours) might be related to river regulations. In the case that such regulations are present, hydrological process modelling would require the deductive explanation of changes to be optimal (see e.g., Koutsoyiannis 2011; Montanari and Koutsoyiannis 2014; Koutsoyiannis and Montanari 2015). We extract one 90-year-long time series block from each of the entire time series retrieved (to obtain time series with the same length). Specifically, we keep the first 90 values of each entire time series, since we need time series with no missing values. In fact, missing values are more often in more recent data.

From the total of the 90-year-long time series, we finally keep only those that simultaneously do not contain missing values and have resulted from the aggregation of daily time series with missing up to 10/90 × 100% ~ 11% of their values (i.e., 10/90 × 365 ~ 40 daily values per year on average). The retained time series are 599 in number. They originate from stations that are mostly located in North America (Region A) and Europe (Region B), as presented in Figure 1. These 599 river flow time series are exploited for both big data forecasting and computation of selected river flow statistics. The related methodological information is detailed in Sections 2.2 and 2.3, respectively. Catchment attribute information is provided in Appendix B.

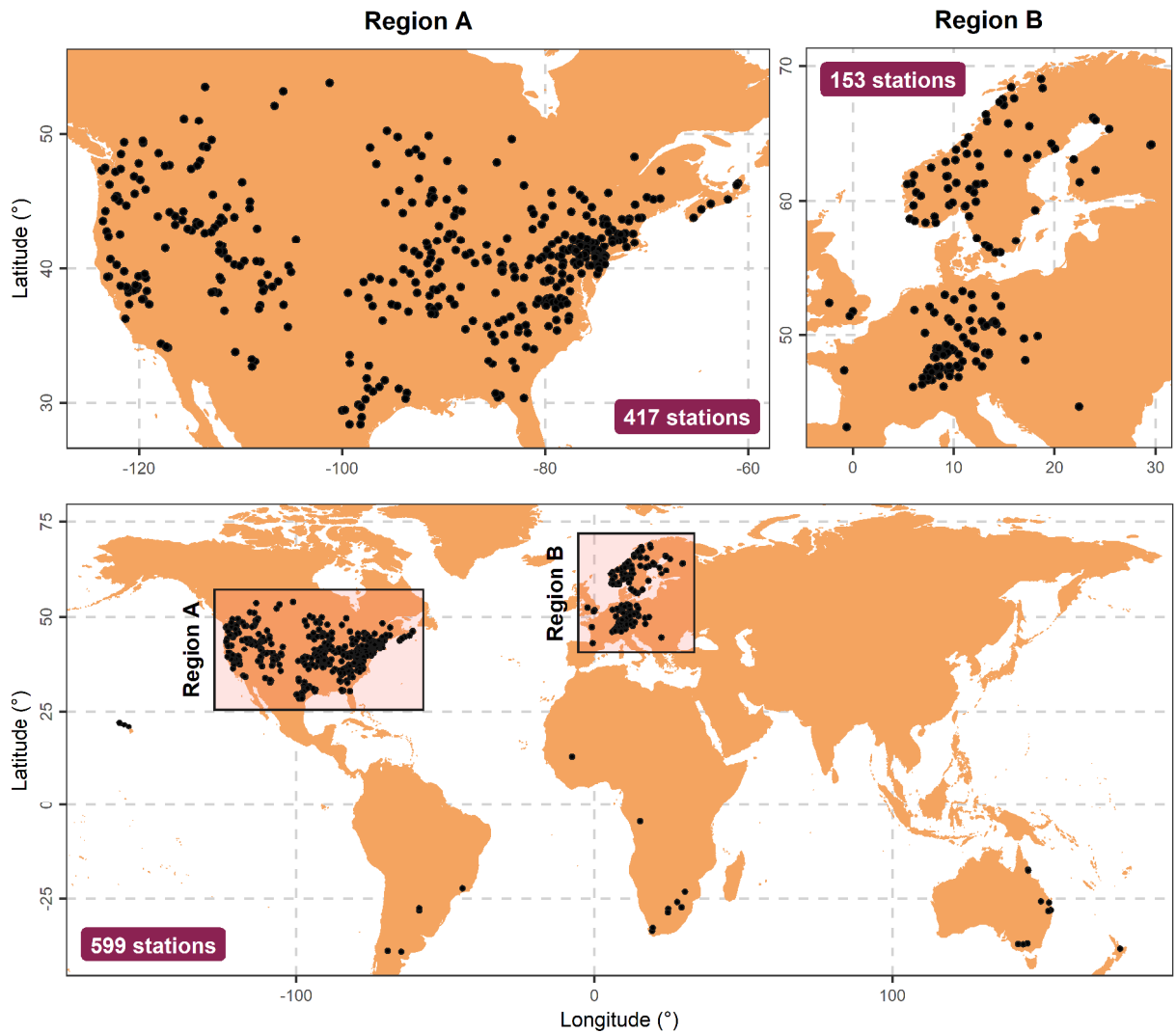

Figure 1. Maps of the locations of the river flow stations exploited in the study, and definition of Regions A and B. The data have been sourced from Do et al. (2018b) using information from Gudmundsson et al. (2018b).

## 2.2 Forecasting and testing methodology

### 2.2.1 Base methods

We implement five individual forecasting methods (hereafter referred to as "base methods") with different characteristics and properties. The first and simplest one is Naïve. Its forecasts are equal to the last value of the training segment (i.e., the time series segment containing all historical information exploited in forecasting). Because of its simplicity, interpretability and good performance in various real-world forecasting problems, this method is frequently used as a benchmark in the literature (see e.g., Hyndman and Athanasopoulos 2018, Section 3.1). It is also our benchmark herein.

Another (rather) simple method that is implemented in this study is SES. As indicated by its name, SES is the simplest method from the exponential smoothing family of forecasting models. This method calculates weighted averages by assigning weights to the historical values. These weights decrease exponentially as we move from the most recent historical values to the most distant ones. As explained in Hyndman and Athanasopoulos (2018, Section 7.1), SES is in-between our last-observation benchmark (see above) and the average method. In fact, these two simple methods can also be considered to calculate weighted averages. In Naïve, a weight equal to 1 is assigned to the last value of the training segment and weights equal to 0 are assigned to the remaining values. In the average method, on the other hand, equal weights (summing up to 1) are assigned to all values of the training segment. The one-step-ahead forecast of SES at time $t + 1$ is a weighted average of the last value of the training segment $x_t$ and its forecast $f_t$, which is equivalent to a weighted average of all observations of the training segment $x_1, …, x_t$. The forecasting equation includes two parameters, which are estimated on the training segment by using the maximum likelihood method. SES performed well on the M3-competition data (Hyndman et al. 2002).

The second exponential smoothing method exploited in the study is CES. CES is a non-linear forecasting method that uses the theory of functions of complex variables (i.e., variables involving a real part and an imaginary part). Unlike most exponential smoothing models, this method does not perform time series decomposition, thereby avoiding the arbitrary distinction between level and trend components, and eliminating the model selection procedure (Svetunkov and Kourentzes 2016). A simple combination of CES and three other forecasting methods (specifically, the exponential smoothing ETS, automatic ARIMA and dynamic optimised theta methods) performed well in the M4-competition (Petropoulos and Svetunkov 2020).

We also implement automatic ARFIMA. The training procedure of this forecasting method is detailed in Hyndman et al. (2019; see also Hyndman and Khandakar 2008; Fraley et al. 2012). In summary, the following sequential steps are taken: (a) Estimation of $d$ by fitting an ARFIMA(2,$d$,0) model; (b) fractional differencing of the time series by using the $d$ estimate obtained at step (a); (c) selection of an ARMA model for the fractionally differenced time series obtained at step (b) by using the maximum likelihood method; and (d) re-estimation of the full ARFIMA($p$,$d$,$q$) model by using the algorithm by Haslett and Raftery (1989). For the definition and theory of ARFIMA models, the reader is referred to Wei (2006, pp. 489–494).

The last base method implemented in the study is Prophet, a fast and quite interpretable machine learning method. This method was originally designed for forecasting time series that are of interest in Facebook. Details on Prophet are available in Taylor and Letham (2018, Section 3). In summary, it uses the additive decomposable time series model by Harvey and Peters (1990), which is similar to the generalized additive model by Hastie and Tibshirani (1987). Prophet is considered to be more flexible than AR(F)IMA models. For an explanation, see Taylor and Letham (2018, Section 3).

### 2.2.2 Simple combination methods

To combine the forecasts of the base methods, we employ a simple forecast combiner. This combiner computes the median of forecasts and has not been used in hydrology before. We apply 26 variants of this combiner (also referred to as "simple combination methods"). These variants are defined by all the possible combinations (per two, three, four or five) of the base methods; therefore, each of them exploits a number of forecasts that is varying from two to five. In detail, two forecasts are exploited by 10 variants, which simply compute the sample mean of these forecasts. (The median of two forecasts is equal to their mean). Three forecasts are exploited by 10 variants, which simply compute the sample median of these forecasts. Four forecasts are exploited by five variants, which simply compute the sample mean of the two middle (in magnitude) forecasts. (The median of four forecasts is equal to the mean of the two middle forecasts). Lastly, five forecasts are exploited by one variant, which simply computes the sample median of these forecasts. When more than two forecasts are exploited, the median combiner of forecasts is theoretically expected to be more robust than simple averaging (delivering a mean value). The simple combination methods of this work can be perceived as variants of a flexible hydrological time series forecasting methodology. This new methodology is formulated in Figure 2, with which aim (1) of the study (see Section 1.3) is addressed.

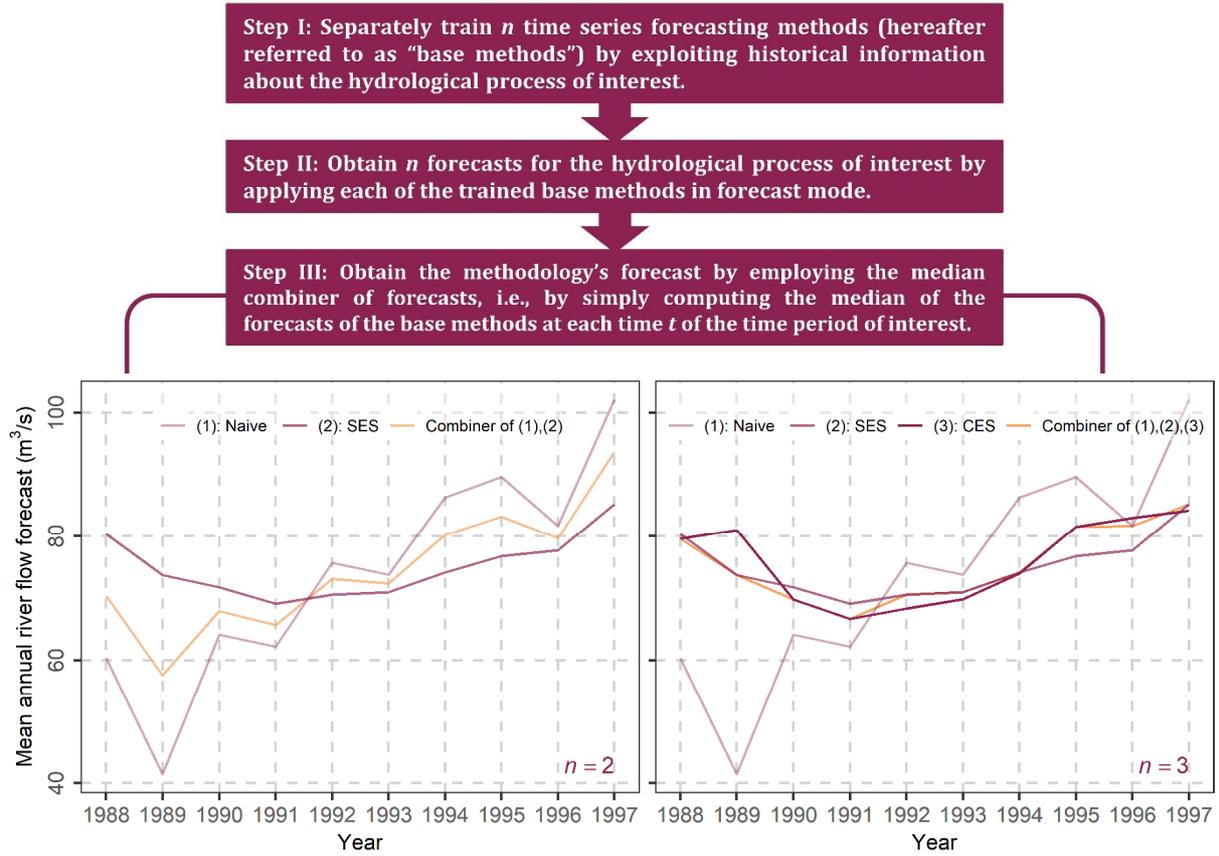

Figure 2. Schematic summarizing the three consecutive steps adopted within the flexible hydrological time series forecasting methodology proposed in the study (upper panel). The number of time series forecasting methods exploited by the methodology ($n$) should be equal or larger than two. Step III is illustrated in two time series plots (bottom panels). The illustration is made for the case in which two time series forecasting methods are utilized by the methodology (left plot) and for the case in which three time series forecasting methods are utilized by the methodology (right plot). For these plots, mean annual river flow information for the Saint Louis River, United States is exploited (as implied by the methodological framework of the study).

### 2.2.3 Forecast accuracy metrics

We compute five forecast accuracy metrics. For the definitions of these metrics, let us consider a forecasted time series of $N$ values and its target values, with the former denoted with $f_1, f_2, ..., f_N$ and the latter with $x_1, x_2, ..., x_N$.

The mean absolute error (MAE) metric is defined by

$$\text{MAE} := (1/N) \sum_N |f_i - x_i| \tag{1}$$

The mean absolute percentage error (MAPE) metric is defined by

$$\text{MAPE} := (1/N) \sum_N |100(f_i - x_i)/x_i| \tag{2}$$

The median absolute error (MdAE) metric is defined by

$$\text{MdAE} := \text{median}_N \{|f_i - x_i|\} \tag{3}$$

where median$_N${} denotes the sample median of $N$ data points.

The median absolute percentage error (MdAPE) metric is defined by

$$\text{MdAPE} := \text{median}_N\{|100(f_i - x_i)/x_i|\} \tag{4}$$

The root mean square error (RMSE) metric is defined by

$$\text{RMSE} := ((1/N) \sum_N (f_i - x_i)^2)^{1/2} \tag{5}$$

MAE, MdAE and RMSE are scale-dependent, while MAPE and MdAPE are scale-independent. Discussions and justifications on the appropriateness of these metrics for forecasting performance assessment can be found in Armstrong and Collopy (1992), and Hyndman and Koehler (2006).

### 2.2.4 Forecasting and testing workflow

We describe here below the forecasting and testing workflow for a single 90-year-long river flow time series; the extension to all time series is straightforward.

Step 1. Extract 10 data blocks from the time series. These blocks correspond to 80-year-long time periods and serve as training segments for the application of the base methods. They start from the 1st, 2nd, 3rd, ..., 10th value of the time series and end with its 80th, 81st, 82nd, ..., 89th value, respectively.

Step 2. Train the base methods (see Section 2.2.1) separately on each training segment. In total, 5 (number of base methods) × 10 (number of training segments) = 50 trained models are obtained.

Step 3. Obtain the one-year ahead forecasts of the base methods for the last 10 mean annual river flow values by using all the trained models obtained at step 2 in one-step ahead forecast mode. In total, 50 one-year ahead forecasts are obtained. These forecasts form five 10-point-long forecasted time series (corresponding to the last 10 mean annual river flow values), each obtained by using a different base method. Negative forecasted values are replaced with zero.

Step 4. Compute the one-year ahead forecasts of the simple combination methods (see Section 2.2.2) by exploiting the forecasted time series obtained at step 3. In total, 26 10-point-long forecasted time series are obtained, as many as the simple combination methods. Examples of one-year ahead forecasts delivered by the simple combination methods, in comparison to the one-year ahead forecasts delivered by the base methods and the target values, are presented in Figure 3.

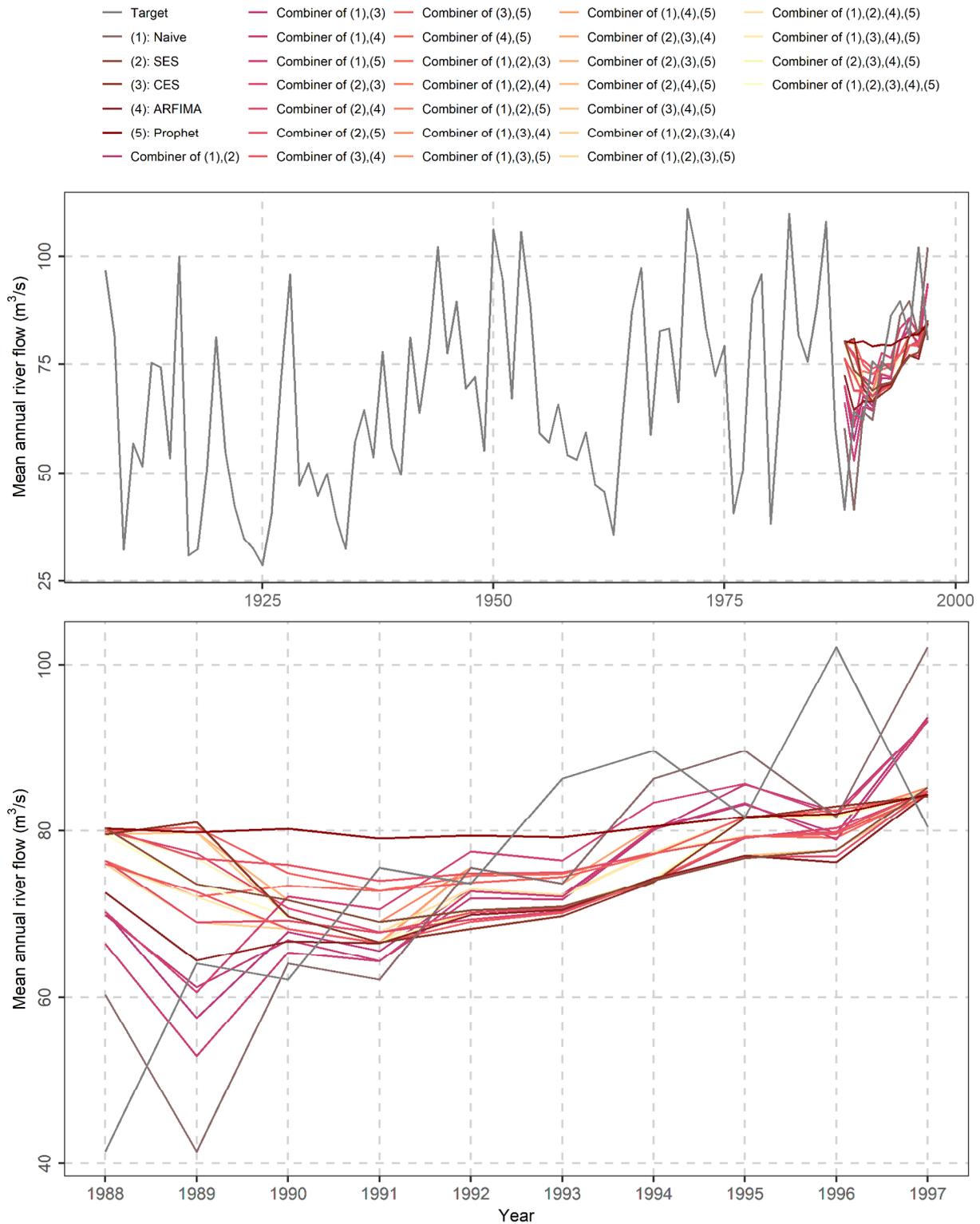

**Figure 3.** Examples of one-year ahead forecasts on the mean annual river flow time series from the Saint Louis River, United States.

Step 5. Compute the metric values (see Section 2.2.3) for all 10-year-long forecasted time series. In total, 5 (number of metrics) × 31 (number of 10-point-long forecasted time series, obtained at steps 3 and 4) = 155 metrics values are computed.

Step 6. For each different metric, rank the methods from 1st (best-performing) to 31st (worst-performing) according to the metric values. In total, five rankings are performed, as many as the different metrics computed.

Step 7. For each different metric, compute the relative improvements provided by all methods with respect to the benchmark (i.e., Naïve). These relative improvements are given by Equation (6). In this Equation, RI$_{m,M}$ denotes the relative improvement in terms of metric $m$ obtained by using the forecast of method $M$ instead of using the forecast of Naïve, while $m_M$ and $m_{\text{Naïve}}$ denote the values of metric $m$ computed for the forecasts of the methods $M$ and Naïve, respectively. In total, 3 (number of scale-dependent metrics; see Section 2.2.3) × 30 (number of all methods excluding the benchmark) = 90 relative improvements are computed.

$$\text{RI}_{m,M} := (m_{\text{Naïve}} - m_M)/m_{\text{Naïve}} \qquad (6)$$

To summarize the results obtained through steps 1–7, for each pair {method, metric} we compute the (i) method's average ranking (for all metrics), (ii) method's average metric value (for the scale-independent metrics), and (iii) method's average relative improvement (for the scale-dependent metrics). The averages are computed across (a) all the examined time series (originating from the Globe), (b) the time series originating from Region A, and (c) the time series originating from Region B.

The results of the forecast performance investigations are presented in Section 3.1 (see also Appendix C), where aims (2)–(4) of the study (see Section 1.3) are addressed. The results in terms of RMSE are also essential for addressing aims (5) and (6) of the study (see Section 1.3); therefore, they are further exploited as detailed in Sections 2.4 and 2.5, respectively.

2.3   Computation of selected river flow statistics

We fit the fractional Gaussian noise process to the entire 90-year long time series of the study (see Section 2.1) by implementing the maximum likelihood method (Hipel and McLeod 1994, Chapter 10.4.3). We use the mean and standard deviation estimates of this stochastic process to compute the maximum likelihood estimates of the coefficient of variation. The latter coefficient is a standardized measure of dispersion. The larger its values, the larger the relative dispersion. We also compute the sample autocorrelation function at lag 1 for the same time series. The sample autocorrelation at lag 1 and the Hurst parameter of the fractional Gaussian noise process are considered as good measures of temporal dependence, specifically of autocorrelation and long-range dependence, respectively. They can take values in the intervals [−1,1] and (0,1), respectively. The larger a positive (negative) sample autocorrelation value at lag 1, the larger the magnitude of the positive (negative) correlation between two subsequent data points in the time series, while Hurst parameter values larger (smaller) than 0.5 indicate long-range dependence (anti-persistence). The magnitude of the long-range dependence is larger for larger Hurst parameter values.

Moreover, we estimate the trend strength and spectral entropy according to Talagala et al. (2018). Specifically, we decompose each original time series into its trend and remainder components by applying Friedman's super smoother (Friedman 1984), and compute the trend strength estimate by using both the estimated variance of the original time series and the estimated variance of the remainder time series. The larger the magnitude of this estimate, the larger the trend strength. For estimating spectral entropy, we exploit the measure by Goerg (2013). According to this latter work, spectral entropy can be used to measure the time series "forecastability". The smaller the spectral entropy values, the larger the "forecastability".

We use the coefficient of variation estimates, the sample autocorrelation at lag 1, the Hurst parameter estimates of the fraction Gaussian noise process, the trend strength estimates and the spectral entropy estimates to characterize the river flow time series. The related results are presented in Section 3.2. These results are exploited as detailed in Section 2.4 for addressing aim (5) of the study (see Section 1.3).

## 2.4 Linear regression analysis

We investigate the existence of possible relationships between predictive performance and statistical characteristics of the river flow time series by exploiting (a) the relative improvements in terms of RMSE provided by all methods with respect to Naïve (see step 7 in Section 2.2.4), (b) the coefficient of variation estimates (see Section 2.3), (c) the estimates of autocorrelation at lag 1 (see Section 2.3), (d) the Hurst parameter estimates (see Section 2.3), (e) the trend strength estimates (see Section 2.3), and (f) the spectral entropy estimates (see Section 2.3). We fit a simple linear regression model between the relative improvements and their corresponding values of each statistic. We also compute the Pearson's *r* coefficient for the five regression settings and present the computed linear correlations. The presentation is made in comparison to the linear correlations between the values of the five statistics. The results of these investigations are presented in Section 3.3.

## 2.5 Predictability assessment

We examine one-year ahead river flow predictability by computing the relative improvements provided by the best-performing method for each of the examined time series (i.e., the method ranked 1st in terms of RMSE) with respect to the last-observation benchmark, and their average across all river flow stations. The larger (smaller) these relative improvements, the larger (smaller) the relative one-year ahead predictability of annual river flow. Zero relative improvements indicate that the maximum levels of one-year ahead river flow predictability are reached by using the last-observation benchmark. The results of these investigations are presented in Section 3.4.

## 3. Results and discussions

## 3.1 Forecasting performance of simple combinations

In this Section, we summarize the results of our big data forecasting experiment. The provided summary is made in a way that facilitates the comparative evaluation of the forecasting methods with respect to their long-run properties. We focus on these properties (and not to every single case itself), since we aspire to support technical applications in a practical way. In fact, the "no free lunch theorem" (Wolpert 1996) implies that we should –by no means– expect to find a method (e.g., a variant of the new methodology) that is always performing better than other methods in solving problems of the same type (e.g., mean annual river flow forecasting problems) in absence of significant information about these problems at hand (see also the related discussions in Papacharalampous et al. 2019a, Section 4.1). Even for a specific catchment, different methods might perform the best for different time periods, while we cannot know in advance which these methods are.

By focusing on the long-run properties (instead of focusing on the case itself), we can effectively facilitate the detection of forecasting methods that are expected to perform better than (or at least comparably to) others on average across a large number of individual cases. Once we have identified the best performing methods (and under which conditions these methods work well) in the long run, we can use these methods either in large-scale forecasting applications to reduce uncertainty to some extent or even in local case studies to increase the chance of obtaining accurate (and, therefore, useful) forecasts. Moreover, forecasting performance patterns revealed by big data forecasting tests could provide possible empirical explanations on how forecasting methods work. Such explanations are important, since they can build confidence in using the methods.

In light of the above, we here focus on answering the following research questions:

(i) Could the new methodology be useful in reducing uncertainty within long-run hydrological forecasting applications and increasing the chance of obtaining accurate forecasts in local case studies?

(ii) Under which conditions the performance of the methodology is maximized?

(iii) Can we identify any interesting forecasting performance patterns?

To support the below-provided summary and answers to these questions, we present: (a) the average rankings of the 31 methods conditional on the metric for the entire dataset (composed by 599 time series; see the presentation under the label "Globe"), Region A and Region B (see Figure 4); (b) the average relative improvements provided by all methods with respect to the benchmark conditional on the metric and the region (see Figure 5); (c) the number of times that each method is ranked in the first five, ten and fifteen places conditional on the metric (see Figure 6); and (d) the rankings of the methods in terms of RMSE conditional on the method and the river station (see Figure 7). Additional visualizations can be found in Appendix C, along with a more detailed presentation of the results of our comparative investigations. Visualizations focusing on the case itself can also be found in the same Appendix. These latter visualizations could be particular important for the specific rivers examined in this work.

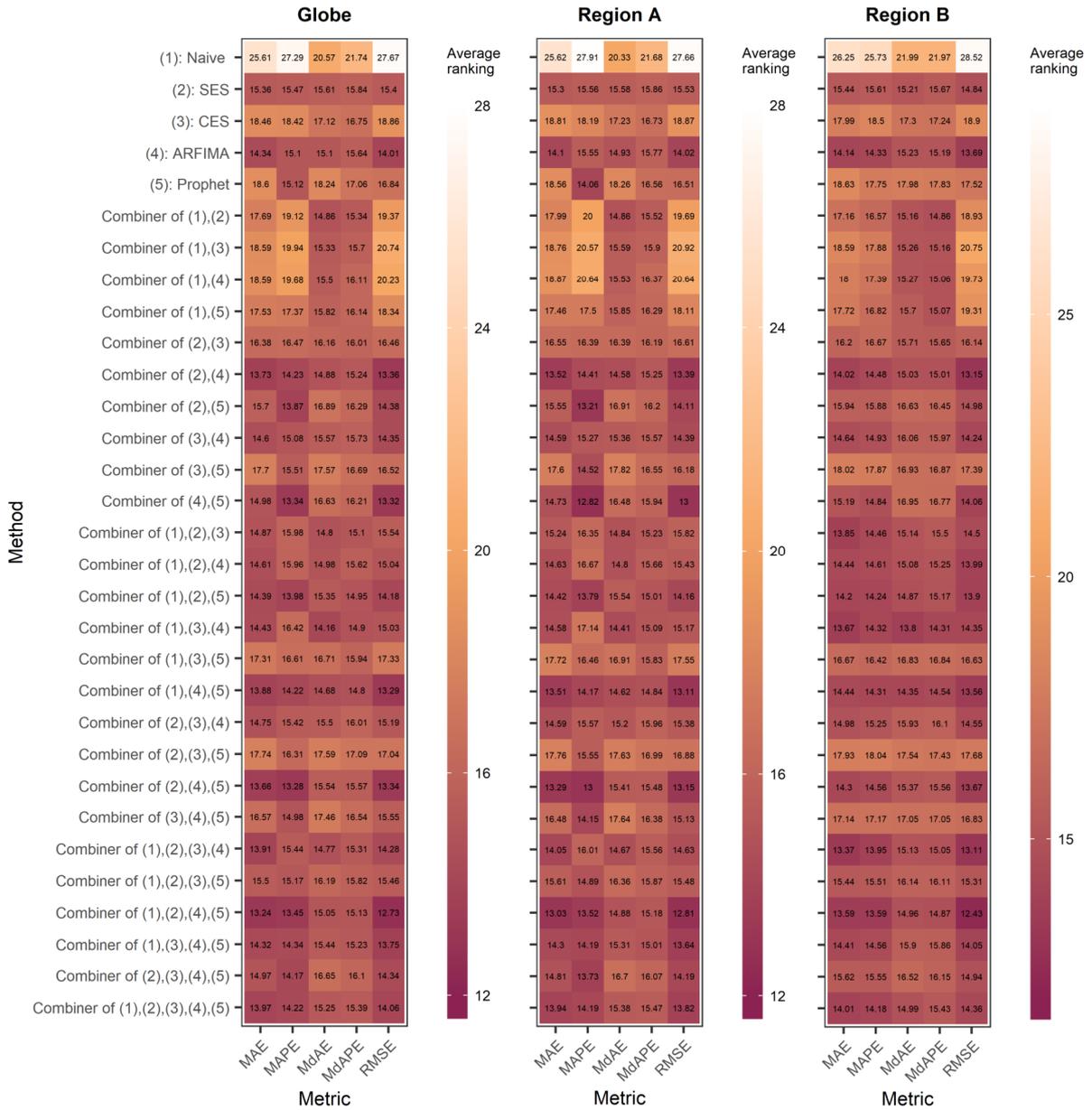

Figure 4. Average rankings of the methods conditional on the region and the metric. Each presented value concerning the entire dataset, Region A and Region B summarizes 599, 417 and 153 values, respectively. The lower the average ranking the better the average-case performance.

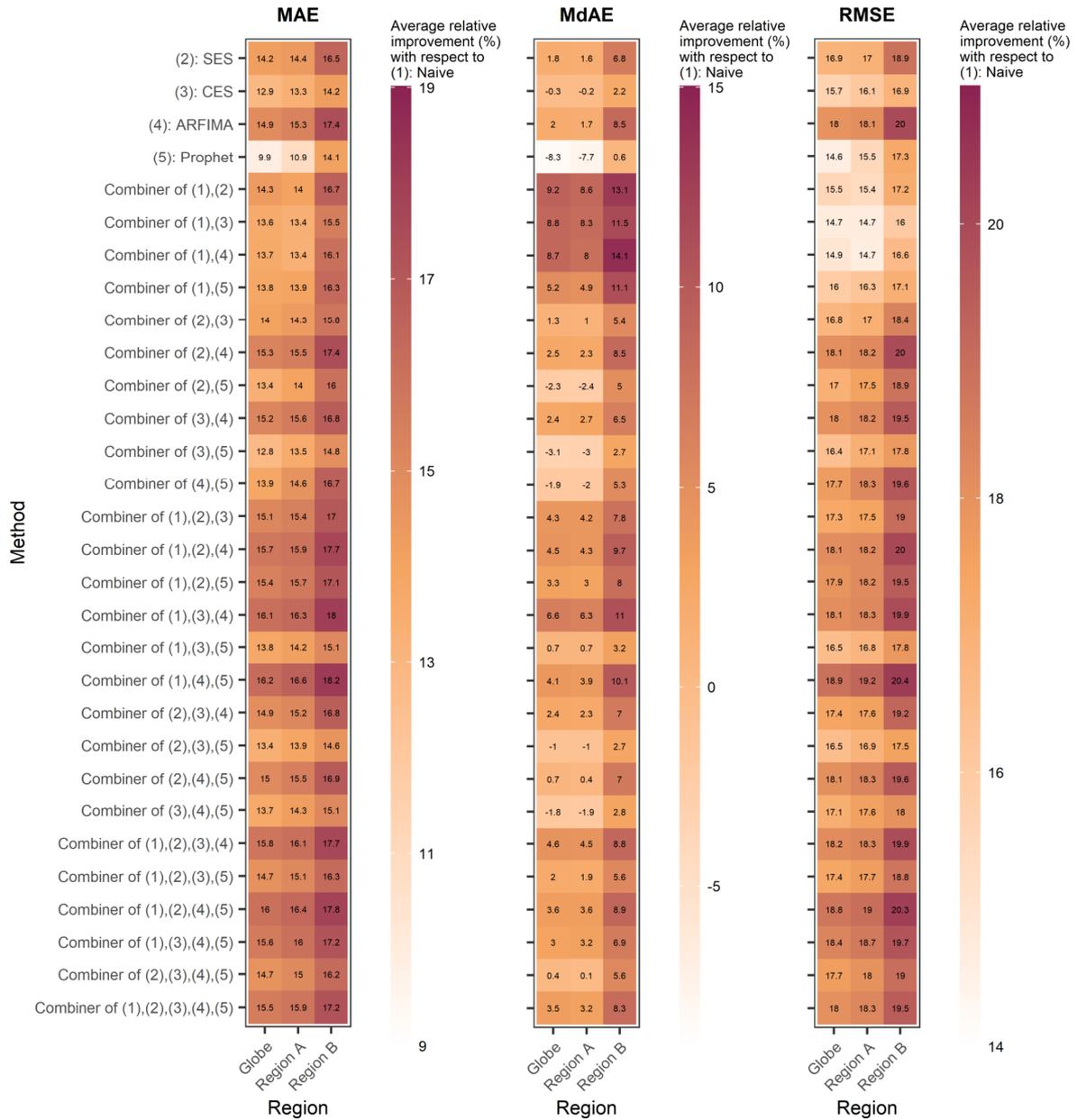

Figure 5. Average relative improvements provided by all methods with respect to the benchmark conditional on the metric and the region. Each presented value concerning the entire dataset, Region A and Region B summarizes 599, 417 and 153 values, respectively.

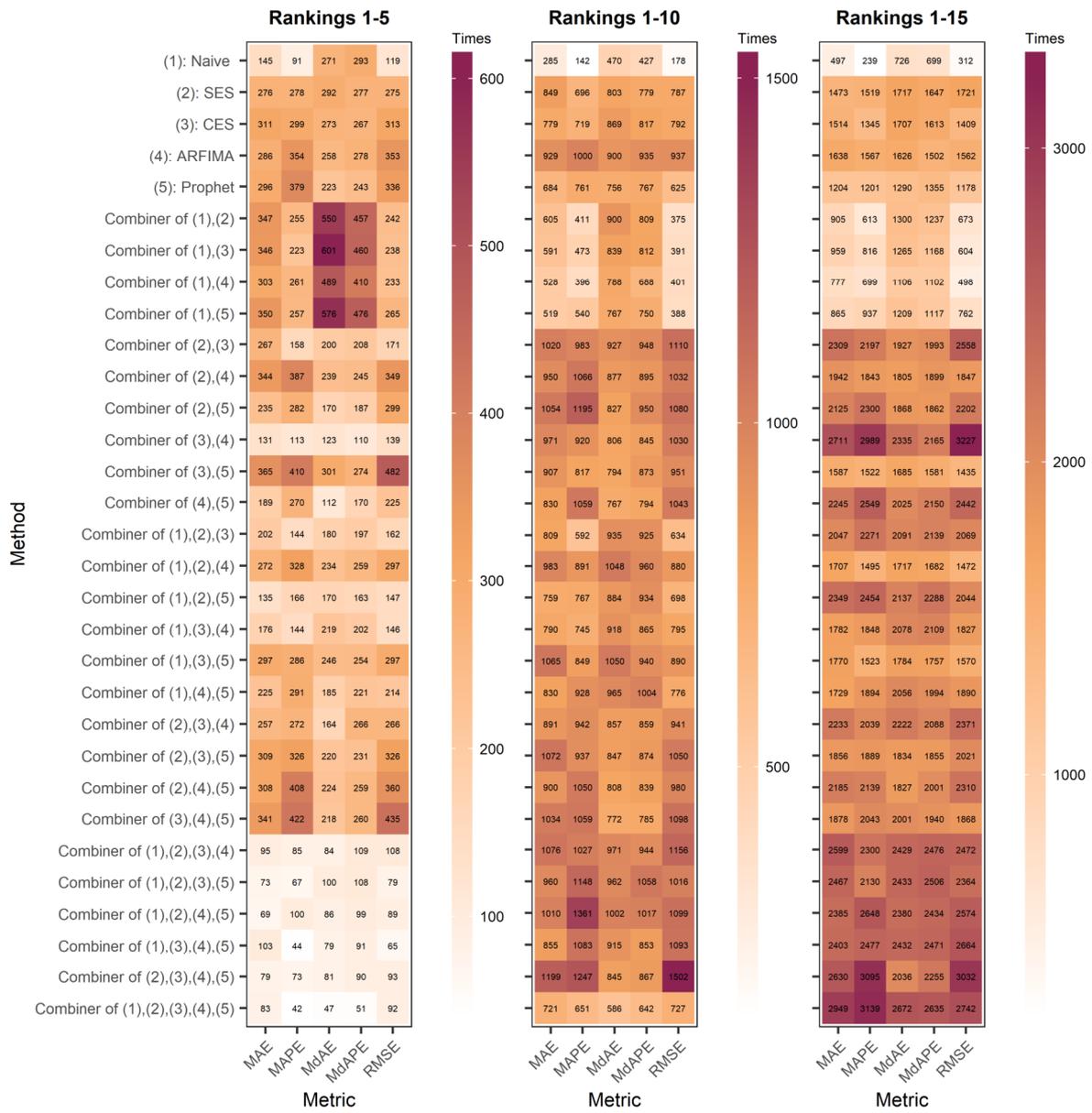

Figure 6. Times that each method is ranked among the five, ten and fifteen best-performing ones in all conducted tests conditional on the metric. The lower the ranking the better the performance.

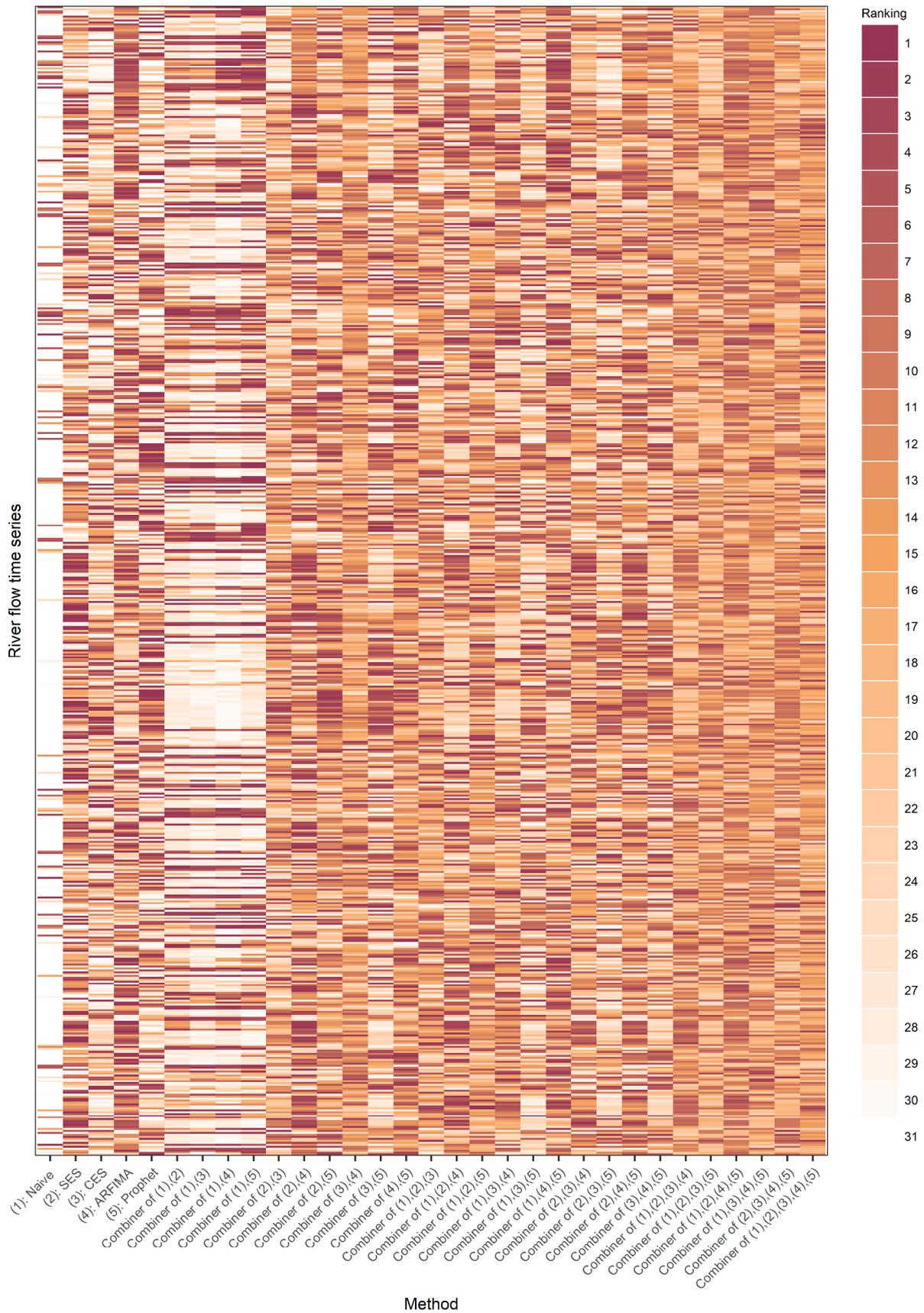

Figure 7. Rankings of the methods in all conducted tests in terms of RMSE. The lower the ranking the better the performance.

The most important findings derived from our comparative tests are the following:

(A) Independently of the region and the metric, most of the variants of the new methodology exhibit average-case rankings that are at least as good as or even better than the average-case rankings of their base methods (see Figure 4).

(B) In terms of average relative improvements (see Figure 5), all four variants that exploit only Naïve and another base method improve over the performance of Naïve from approximately 13% to approximately 17%. The remaining variants are at least as good (roughly), and usually better, than each of the base methods.

(C) On average across all the examined river flow stations, the best-performing variant (exploiting Naïve, ARFIMA and Prophet as base methods) has improved over the performance of the benchmark by 18.9% in terms of RMSE. The respective average relative improvements provided by ARFIMA, SES, CES and Prophet have been found equal to 18.0%, 16.9%, 15.7% and 14.6%, respectively (see Figure 5).

(D) Based on points (A)–(C) above, we believe that the new methodology could be useful in reducing uncertainty within long-run hydrological forecasting applications and increasing the chance of obtaining accurate forecasts in local case studies.

(E) Despite being the first and second worst-performing methods in the long run in terms of RMSE, Naïve and Prophet helped in obtaining the best average-case improvement (in the same terms) for the entire dataset. This is an interesting –yet not interpretable– outcome highlighting the fact that in data-driven frameworks results are not (and should not be expected to be) explainable always.

(F) In all the cases that the worst-performing base method (i.e., Naïve) is exploited by the new methodology, the strategy of exploiting only two base methods results in larger average-case rankings and smaller average relative improvements than the strategy of combining more than two base methods (see Figures 4 and 5). Therefore, for maximizing the benefits of the new methodology, its use is suggested with more than two base methods (given the fact that we could –by no means– know in advance which method will be bad-performing for the case study of our interest).

(G) Lastly, the best average-case performances seem to be achieved by those methods mostly exhibiting medium rankings in the examined individual cases, and not by those mostly ranked either in the first or in the last positions depending on the examined individual case (see Figures 6 and 7 in comparison to Figures 4 and 5).

We believe that the illustrations presented in this subsection and the delivered interpretations and insights (see also Appendix C) should be encountered as an empirical explanation of how simple combination methods in general, and the median combiner of forecasts in particular, could manage to reduce uncertainty in various geoscientific modelling contexts. In the problem examined herein (i.e., one-step ahead forecasting of annual river flow time series), considerable improvements seem to be achieved in the long run by sacrificing some excellent yet case-dependent forecasting performances.

Before moving to the remaining investigations and discussions (that focus less on the new methodology and more on the river flow processes), it is highly relevant to consider the fact that ARFIMA and SES (i.e., the best-performing individual methods in our experiments) have been identified as two hard-to-beat (in the long run) time series forecasting methods within big data time series forecasting competitions (conducted in the forecasting field). Here, we have managed to beat both these well-performing methods (that could excellently serve as good benchmarks in hydrological time series forecasting tests) by applying several variants of the simple combination methodology of the study. This is fairly one of the most important outcomes of this study.

## 3.2 Mean annual river flow characterization

Summaries of the computed hydrological statistics are presented in Figures 8 and 9. We observe that the coefficient of variation estimates exhibit a mean value approximately equal to 0.4, and that their histogram is right-skewed (with more than half of the estimates being lower than the mean). Autocorrelation at lag 1 is mostly positive with a mean approximately equal to 0.2, indicating low correlation on average between two subsequent data points in the river flow time series. Moreover, the Hurst parameter is mostly larger than 0.5 with a mean slightly larger than 0.6, indicating mostly low (but not negligible) long-range dependence on average. The trend strength estimates have a mean value approximately equal to 0.2 and a histogram that is right-skewed, indicating that the trend strength is mostly low for the examined dataset. Lastly, the spectral entropy estimates are mostly large (with a mean value approximately equal to 0.97 and a left-skewed histogram), indicating low "forecastability" according to Goerg (2013). This latter characterization highlights even more the significance of the improvements achieved by using the new methodology (see Section 3.1). It might also mean that these improvements would be larger for processes with larger "forecastability" (e.g., seasonal or monthly processes).

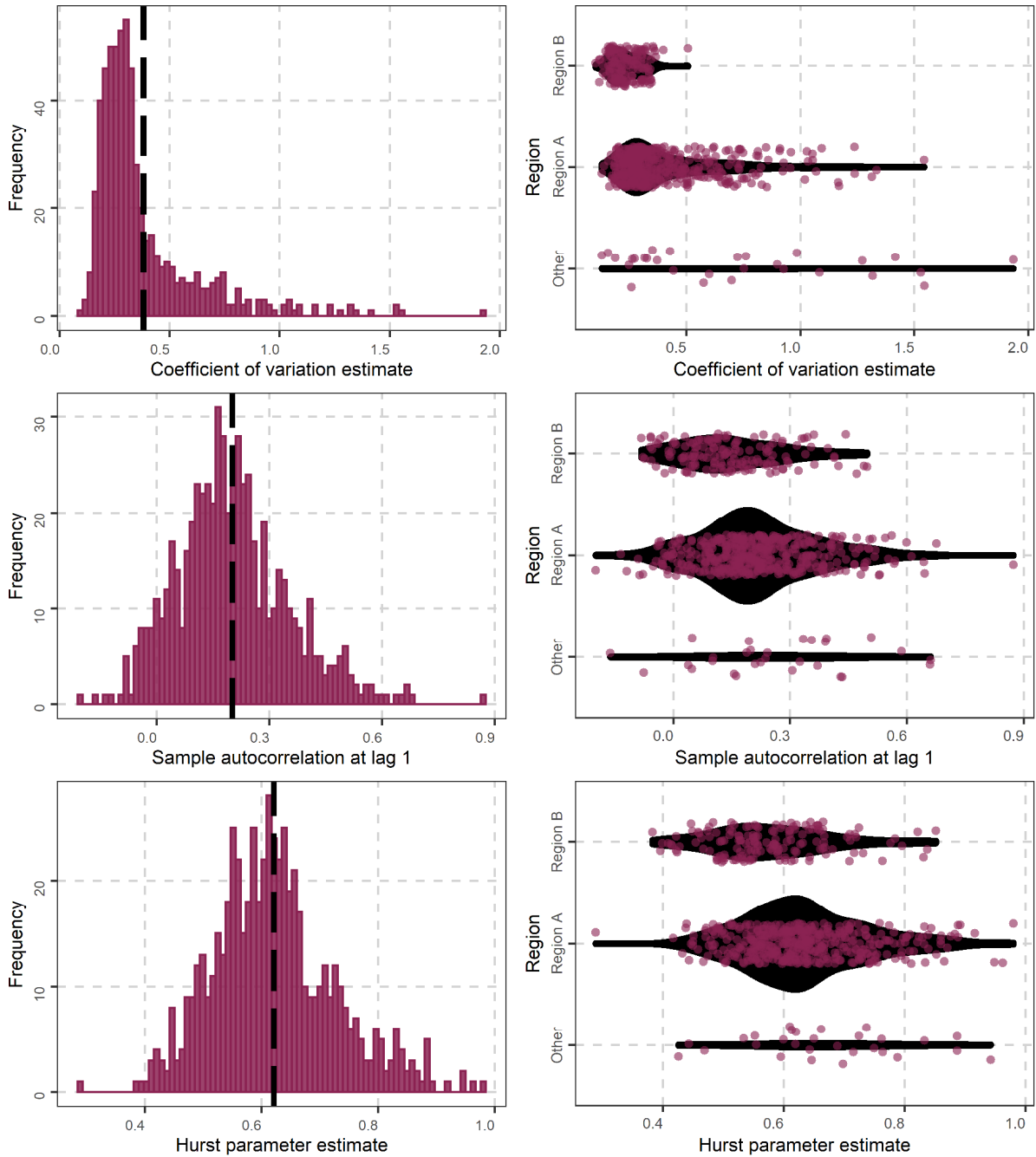

Figure 8. Coefficient of variation estimates (top panels), estimates of autocorrelation at lag 1 (middle panels) and Hurst parameter estimates of the fractional Gaussian noise process (bottom panels) for the river flow time series of the study. The black dashed lines in the histograms (left panels) denote the mean values of the summarized estimates. Each histogram summarizes 599 values. In the side-by-side violin plots (right panels), the estimates are presented as jitter points. Each violin (denoted with black colour) summarizes –in terms of sample density– 417, 153 and 29 values for Region A, Region B and other regions, respectively. Violin areas are scaled proportionally to the number of estimates.

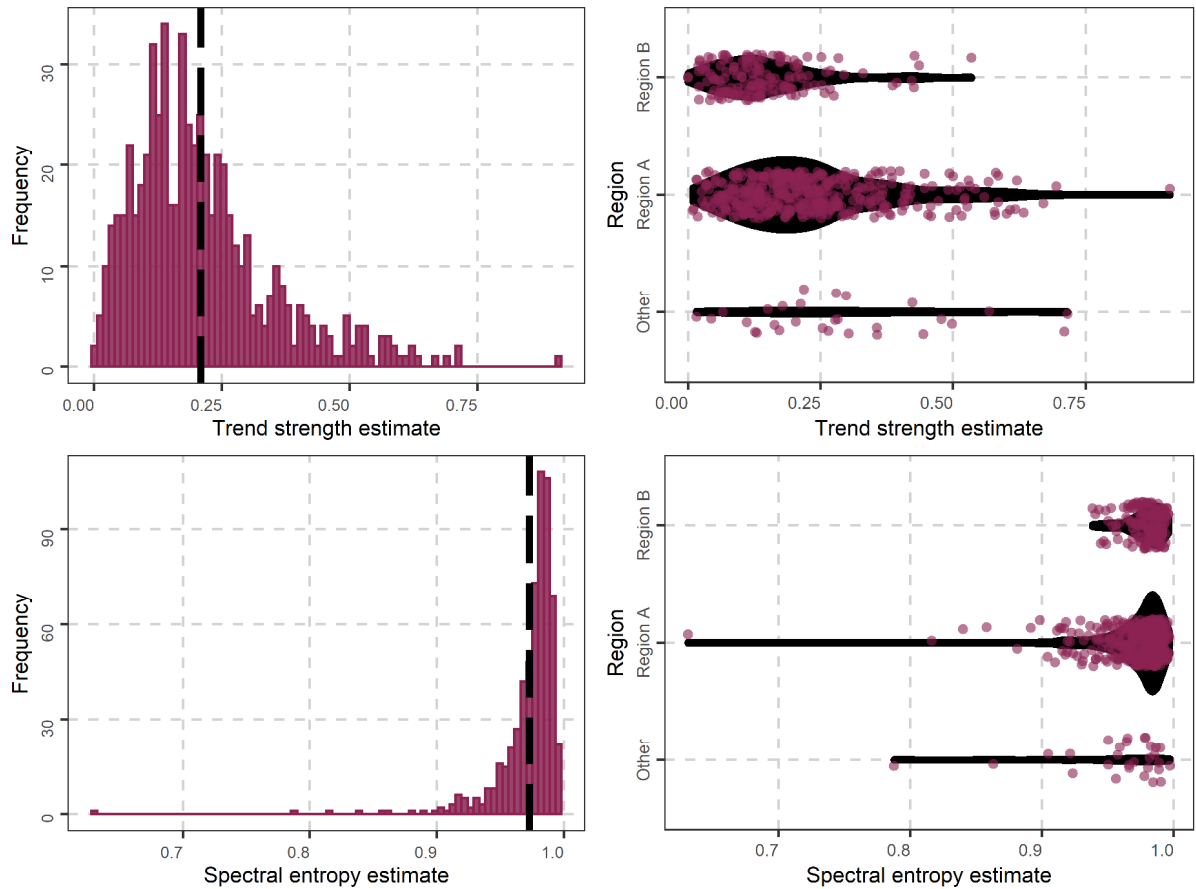

Figure 9. Trend strength estimates (upper panels) and spectral entropy estimates (bottom panels) of the river flow time series of the study. The black dashed lines in the histograms (left panels) denote the mean values of the summarized estimates. Each histogram summarizes 599 values. In the side-by-side violin plots (right panels), the estimates are presented as jitter points. Each violin (denoted with black colour) summarizes –in terms of sample density– 417, 153 and 29 values for Region A, Region B and other regions, respectively. Violin areas are scaled proportionally to the number of estimates.

Conducting hydrological analyses, e.g., for studying temporal dependence in hydrological processes or for analysing trends, is of traditional interest to hydrological scientists (see e.g., Montanari et al. 1997, 1999, 2000; Koutsoyiannis 2002; Montanari 2012; Markonis et al. 2018; Tyralis et al. 2018). Therefore, exploiting the big dataset of this study for better understanding and characterizing river flow processes is by itself an important outcome from a theoretical point of view with several implications in practice (e.g., within stochastic simulation frameworks). Yet, the results of the mean annual river flow characterization are mostly auxiliary herein for delivering the results of Section 3.3.

3.3  Forecasting performance versus river flow statistics

To better understand forecasting performance, but also to explore the possibility of benefiting from case-informed integrations of diverse time series forecasting methods within systematic frameworks in hydrology, with Figures 10–12 we examine how the relative improvements provided by all methods with respect to Naïve change with increasing relative dispersion, autocorrelation, long-term persistence, trend strength and "forecastability" of the river flow processes. We find only loose (but not negligible) relationships. These relationships imply that the last year's observation is more likely to consist a better forecast than the forecast produced by sophisticated methods as the magnitudes of relative dispersion, autocorrelation, long-term persistence, trend strength and "forecastability" increase. Analogous investigations conducted between predictive performance and several catchment attributes (see Appendix D) result in even less pronounced patterns, an outcome that could perhaps be attributed to the examined time scale (i.e., the annual one). Although all identified relationships are loose (as also highlighted by their direct comparison to other relationships in Figure 13), we believe that their exploitation within properly designed frameworks could result in forecasting performance improvements. Research should, therefore, focus on the development of such frameworks to allow case-informed integrations of diverse time series forecasting methods in hydrology.

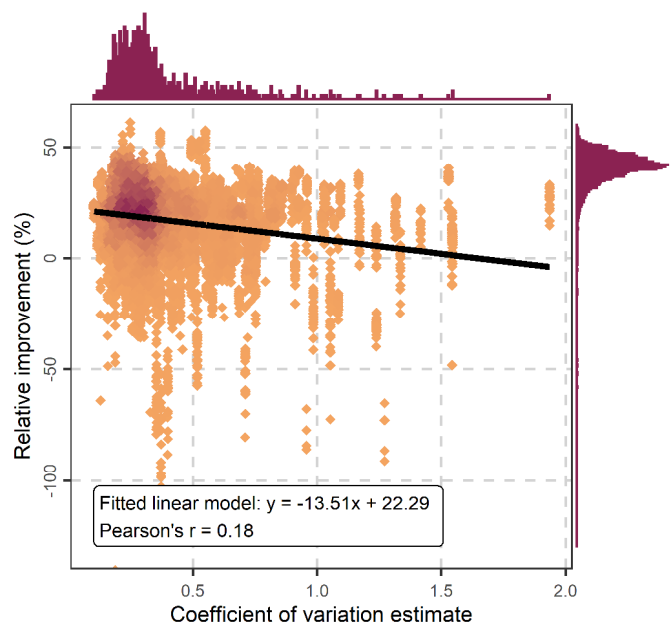

Figure 10. Aggregated relative improvements in terms of RMSE provided by each method with respect to the benchmark for each of the 599 catchments (30 × 599 = 17 970 values) in comparison to the coefficient of variation estimates. Orange and magenta data points denote low and high density, respectively, while the black line denotes the linear model fitted to these data points. The vertical axis has been truncated at −130.

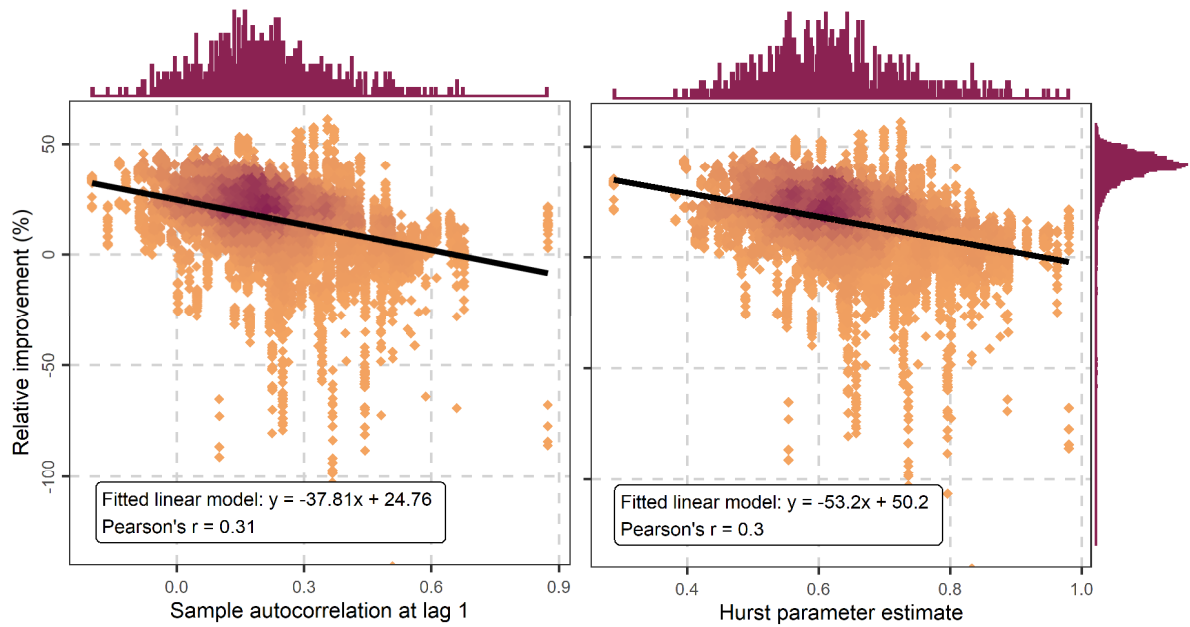

Figure 11. Aggregated relative improvements in terms of RMSE provided by each method with respect to the benchmark for each of the 599 catchments (30 × 599 = 17 970 values) in comparison to the values of sample autocorrelation at lag 1 (left panel) and the Hurst parameter estimates of the fractional Gaussian noise process (right panel). Orange and magenta data points denote low and high density, respectively, while the black lines denote the linear models fitted to these data points. The vertical axes have been truncated at −130.

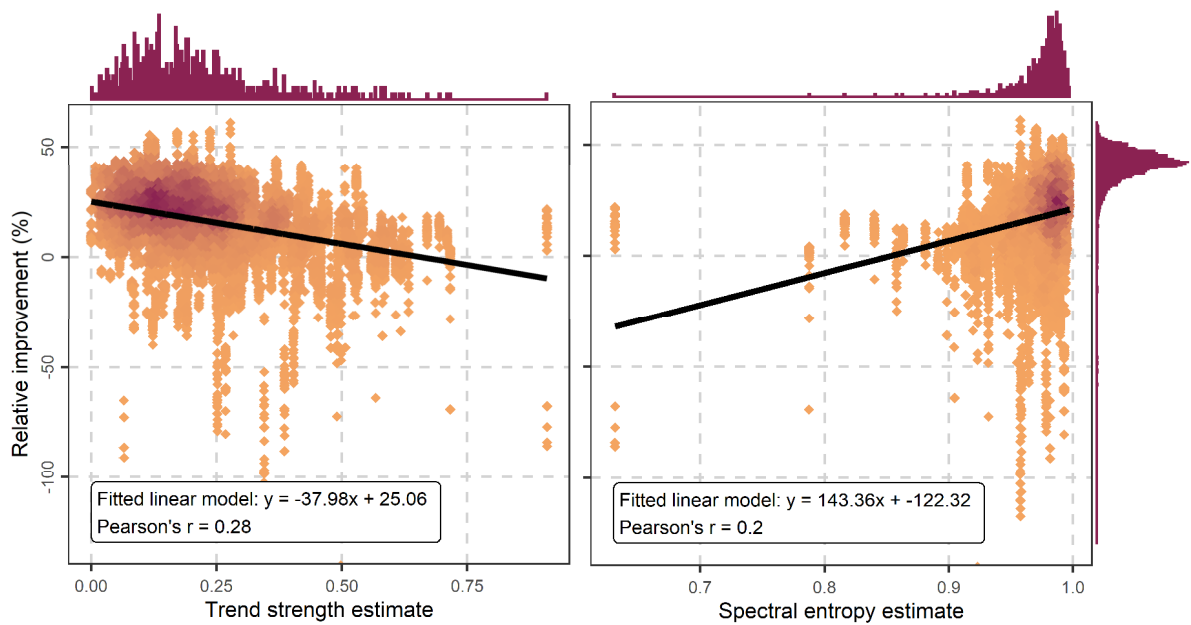

Figure 12. Aggregated relative improvements in terms of RMSE provided by each method with respect to the benchmark for each of the 599 catchments (30 × 599 = 17 970 values) in comparison to the trend strength estimates (left panel) and the spectral entropy estimates (right panel). Orange and magenta data points respectively denote low and high density, while the black lines denote the linear models fitted to these data points. The vertical axes have been truncated at −130.

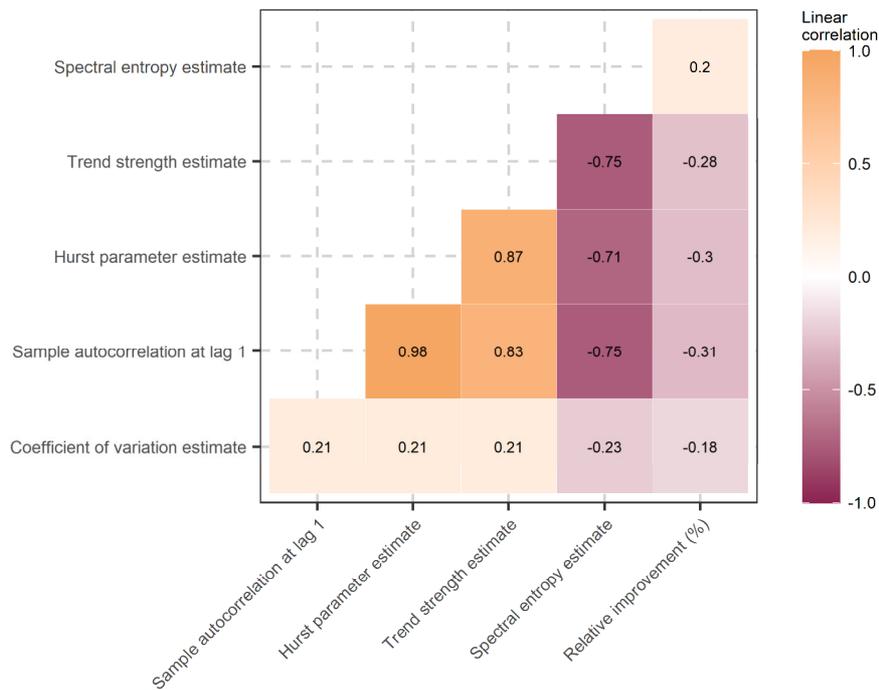

Figure 13. Linear correlations between the computed river flow statistics, as well as between the aggregated relative improvements in terms of RMSE provided by each method with respect to the benchmark for each of the 599 catchments (30 × 599 = 17 970 values) and their corresponding values of river flow statistics.

3.4 One-year ahead river flow predictability

Relative improvements with respect to the benchmark of the study (whose one-year ahead forecast is last year's mean annual river flow) are also considered herein as a proper measure to assess –in relative terms– one-year ahead predictability of river flow and, consequently, as an alternative to the Nash-Sutcliffe efficiency (Nash and Sutcliffe 1970). The latter has been exploited e.g., in Papacharalampous et al. (2018) for assessing multi-step ahead predictability of monthly temperature and precipitation. The good properties of relative error measures are well-understood in the forecasting literature (see e.g., the discussions in Davydenko 2012; Davydenko and Fildes 2013). Moreover, we find that the relative improvements with respect to Naïve can remarkably facilitate interpretability, since they allow answering the following research question: How much better (or worse) are the forecasts of sophisticated methods with respect to simply using the last year's observation?

In Figure 14, we present in an aggregated form the 599 relative improvements in terms of RMSE provided by the best-performing method at the catchment level with respect to the benchmark. The following three observations can be extracted from this figure: (i) The benchmark exhibits the best performance only for seven river flow stations (and is almost as successful as the best-performing method for one river flow station); (ii) the average of the presented relative improvements is approximately 25% (as expected, larger than the average relative improvements in terms of RMSE computed for each method separately; see Figure 5); and (iii) relative improvements that are larger than 40% are infrequent.

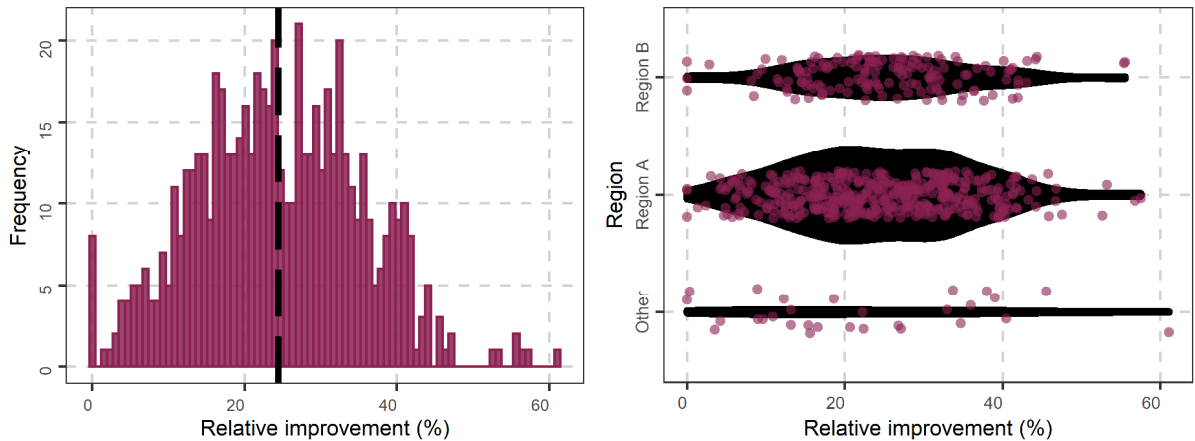

Figure 14. Relative improvements in terms of RMSE provided by the best-performing method (different for each river flow station) with respect to the benchmark for all river flow stations. The black dashed line in the histogram (left panel) denotes the average of the presented relative improvements. The histogram summarizes 599 values. In the side-by-side violin plots (right panel), the estimates are presented as jitter points. Each violin (denoted with black colour) summarizes –in terms of sample density– 417, 153 and 29 values for Region A, Region B and other regions, respectively. Violin areas are scaled proportionally to the number of estimates.

Lastly, the degree of relative one-year ahead river flow predictability characterizing each of the examined river flow processes is also of interest. Therefore, in Figure 15 we present the relative improvements in terms of RMSE provided by the best performing method at the catchment level with respect to the benchmark in Regions A and B. We observe that the 417 North American river flow stations investigated in the study cannot be grouped into sub-regions with similar one-year ahead river flow predictability. The same holds for the 153 European river flow stations of the study.

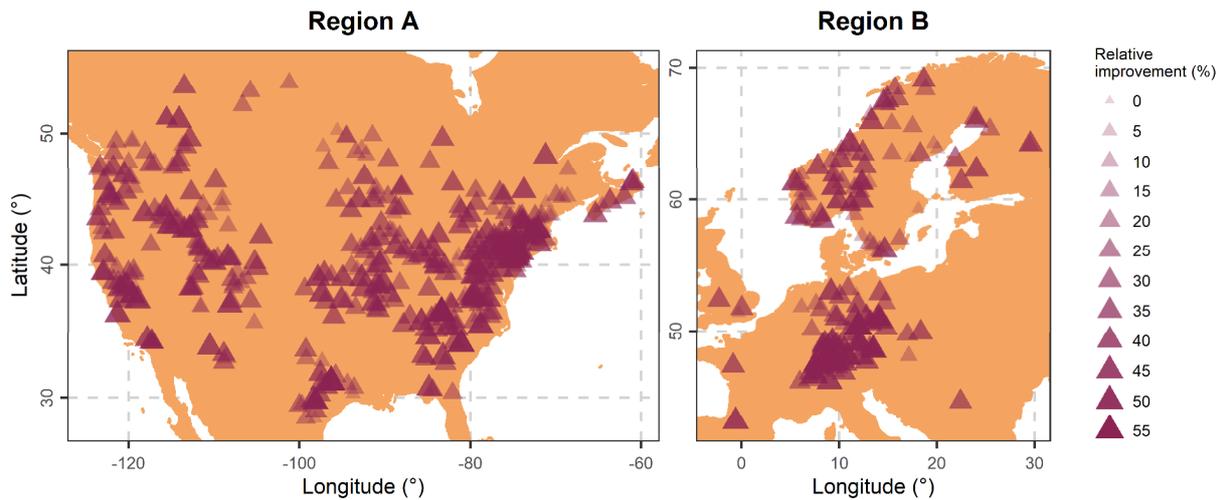

Figure 15. Relative improvements in terms of RMSE provided by the best-performing method (different for each river flow station) with respect to the benchmark for the 417 river flow stations of Region A (left panel) and the 153 river flow stations of Region B (right panel).

## 4. Summary and take-home messages

Big data time series forecasting is rarely performed in the geoscientific and environmental literature. Nonetheless, new forecasting approaches should be first tested on big datasets before applied in engineering contexts. In this work, we have followed this important principle to identify the advantages and disadvantages of a simple and flexible methodology for hydrological time series forecasting. This new methodology is based on the median combiner of forecasts. We have used 90-year-long annual river flow data from 599 river flow stations. These stations are mostly located in North America (417 river flow stations) and Europe (153 river flow stations). We have applied a benchmark scheme (i.e., the one based on the preceding year's river flow), three stochastic models (i.e., simple exponential smoothing – SES, complex exponential smoothing – CES, and autoregressive fractionally integrated moving average – ARFIMA) and a newly proposed (yet well-checked) machine learning algorithm (i.e., Prophet) to forecast the last 10 values of each time series. The application is made in one-step ahead forecasting mode. All the applied individual methods are fully automatic and fast; therefore, they are appropriate for big data time series forecasting. We have also applied 26 variants of the simple combiner. These variants are defined by all the possible combinations (per two, three, four or five) of the five afore-mentioned individual methods. Apart from extensively testing the median combiner of forecasts in the context of one-step ahead annual river flow forecasting, we have also investigated the existence of possible relationships between forecasting performance and five selected statistical characteristics (i.e., relative dispersion, autocorrelation, long-term persistence, trend strength and "forecastability") of the annual river flow time series, and quantified –in relative terms– one-year ahead river flow predictability. Our findings have both practical and theoretical implications.

The key outcomes and take-home messages of this work are the following:

- The simple and flexible hydrological time series forecasting methodology proposed in this work can be used for achieving performance improvements in the long run.
- On average across all the examined river flow stations, the proposed methodology has improved over the performance of the benchmark by 18.9% in terms of root mean square error (RMSE), while the average relative improvements provided by the individual methods (other than the benchmark) have been found equal to 18.0% (ARFIMA), 16.9% (SES), 15.7% (CES) and 14.6% (Prophet).
- Beating ARFIMA and SES (as achieved herein) is particularly important, since these two time series forecasting methods are simultaneously traditional and hard-to-beat in the long run.
- The best-performing method identified in the study (based on the relative improvements in terms of RMSE) is this variant of the proposed methodology that exploits the forecasts delivered by ARFIMA (i.e., the best-performing individual method), the benchmark (i.e., the worst-performing individual method) and Prophet.
- Long-run performance improvements seem to be achieved by sacrificing some excellent yet case-dependent forecasting performances.
- Combining more than two individual methods under the simple combination approach of the study is safer than combining only two individual methods.
- The last-observation benchmark is more likely to outperform sophisticated time series forecasting methods as the magnitudes of relative dispersion, autocorrelation, long-term persistence, trend strength and "forecastability" increase.
- Still, this benchmark outperformed all the remaining 30 methods of this study only for seven river flow time series in terms of RMSE and performed almost as well as the best method for one river flow time series.
- The average of these relative improvements across all the examined stations is approximately 25% in terms of RMSE, i.e., relative one-year ahead river flow predictability is remarkably high.

We would like to conclude by remarking, once again, the high practical relevance of our big data investigations. We believe that simple combination methods in general, and the median combiner of forecasts in particular, could be exploited in various predictive modelling concepts in geoscience and environmental science for increasing robustness and improving predictive performance in the long run. We hope that the present study will increase understanding on how such combinations work and build confidence in their use. We also hope that it will trigger interest in the development of systematic frameworks for case-informed integrations of diverse time series forecasting methods in hydrology.

**Acknowledgements:** We sincerely thank the Editor, the Associate Editor, Dr Bibhuti Bhusan Sahoo and an anonymous referee for their very constructive and fruitful suggestions, which helped us to substantially improve this work.

## Appendix A    Statistical software information

The analyses and visualizations have been performed in R Programming Language (R Core Team 2019). We have used the following contributed R packages: countrycode (Arel-Bundock et al. 2018, 2020), data.table (Dowle and Srinivasan 2019), devtools (Wickham et al. 2019c), dplyr (Wickham et al. 2019b), EnvStats (Millard 2013, 2018), ForeCA (Goerg 2013, 2016), forecast (Hyndman and Khandakar 2008; Hyndman et al. 2019a), fracdiff (Fraley et al. 2012), gdata (Warnes et al. 2017), ggcorrplot (Kassambara A 2019a), ggExtra (Attali and Baker 2019), ggforce (Pedersen 2019), ggplot2 (Wickham 2016a; Wickham et al. 2019a), ggpubr (Kassambara 2019b), gridExtra (Auguie 2017), HKprocess (Tyralis 2016), hydroGOF (Zambrano-Bigiarini 2017), knitr (Xie 2014, 2015, 2019), lubridate (Grolemund and Wickham 2011; Spinu et al. 2020), maps (Brownrigg et al. 2018), MASS (Venables and Ripley 2002; Ripley 2019), matrixStats (Bengtsson 2019), prophet (Taylor and Letham 2018, 2019), plyr (Wickham 2011, 2016b), RColorBrewer (Neuwirth 2014), reshape2 (Wickham 2007, 2017a), rmarkdown (Xie et al. 2018; Allaire et al. 2019), smooth (Svetunkov 2019), tidyr (Wickham and Henry 2019), tidyverse (Wickham 2017b), tsfeatures (Hyndman et al. 2019b) and viridis (Garnier 2018).

**Appendix B   Catchment attribute information**

From Do et al. (2018b), we additionally retrieve selected catchment attribute information, which is particularly important from a hydrological point of view. This information is summarized with Tables B.1–B.9, and exploited in Appendix D. According to Do et al. (2018a), the below-summarized information (provided by Do et al. 2018b) has been retrieved from various large-scale works including the metadata of smaller daily streamflow databases, DEM products (HydroSHEDS and ViewFinder products), the global topographic index dataset by Marthews et al. (2015), the Köppen-Geiger climate classification by Rubel and Kottek (2010; see also Kottek et al. 2006), the Land Cover CCI Climate Research Data Package (CRDP), the Global Lithological Map v1.0 (GLiM) by Hartmann and Moosdorf (2012), the SoilGrids250m dataset by Hengl et al. (2017), the Global reservoir and dam (grand) database Lehner et al. (2011), the Global River Network (GRIN) by Schneider et al. (2017) and the Historical Irrigation Dataset by Siebert et al. (2015).

Table B.1. Summary of the country-based grouping of the catchments (see also Figure 1). The summarized information has been sourced from Do et al. (2018b).

| Country | Number of catchments |
|---|---|
| Argentina | 3 |
| Australia | 9 |
| Brazil | 2 |
| Canada | 43 |
| Congo - Kinshasa | 1 |
| Czechia | 4 |
| Germany | 51 |
| Finland | 7 |
| France | 2 |
| Mali | 1 |
| Netherlands | 1 |
| New Zealand | 1 |
| Norway | 39 |
| Romania | 1 |
| Slovakia | 1 |
| South Africa | 7 |
| Sweden | 20 |
| Switzerland | 24 |
| United Kingdom | 3 |
| United States | 379 |

Table B.2. Summary statistics for topographic characteristics of the examined catchments. The summarized information has been sourced from Do et al. (2018b).

| Variable | Region | Minimum | 1st quartile | Median | Mean | 3rd quartile | Maximum | Sample size |
|---|---|---|---|---|---|---|---|---|
| Station altitude (m) | Globe | -1.80 | 114.93 | 232.34 | 415.90 | 420.51 | 2 717.28 | 499 |
| | Region A | -1.80 | 119.57 | 234.05 | 473.07 | 476.74 | 2 717.28 | 363 |
| | Region B | 0.00 | 63.75 | 189.00 | 246.25 | 373.56 | 1 076.54 | 120 |
| Catchment area (km$^2$) | Globe | 4.7 | 496.0 | 1 880.0 | 40 113.0 | 6 950.0 | 3 475 000.0 | 303 |
| | Region A | 19.4 | 662.8 | 1 945.0 | 13 983.7 | 5 698.0 | 389 000.0 | 136 |
| | Region B | 4.7 | 378.5 | 1 755.0 | 15 609.4 | 9 974.0 | 576 232.0 | 151 |
| Estimated catchment area (km$^2$) | Globe | 0.0 | 44.0 | 1 140.0 | 25 640.0 | 5 714.0 | 3 626 564.0 | 599 |
| | Region A | 0.1 | 1.7 | 1 002.0 | 11 751.0 | 5 025.8 | 542 258.6 | 417 |
| | Region B | 0.0 | 237.9 | 1 059.2 | 14 538.8 | 6 248.2 | 577 998.4 | 153 |
| Mean catchment elevation (m) | Globe | 2.25 | 106.25 | 218.00 | 327.80 | 386.30 | 2 465.86 | 599 |
| | Region A | 2.25 | 96.00 | 182.42 | 338.13 | 372.14 | 2 465.86 | 417 |
| | Region B | 3.50 | 137.50 | 248.30 | 298.40 | 383.10 | 1 210.90 | 153 |
| Maximum catchment elevation (m) | Globe | 9.00 | 413.00 | 795.00 | 1 312.00 | 1 778.00 | 6 793.00 | 599 |
| | Region A | 9.00 | 361.00 | 669.00 | 1 195.00 | 1 646.00 | 4 341.00 | 417 |
| | Region B | 14.00 | 719.00 | 1 181.00 | 1 519.00 | 2 042.00 | 4 571.00 | 153 |
| Mean catchment slope | Globe | 0.00 | 0.32 | 0.87 | 1.35 | 1.83 | 9.60 | 599 |
| | Region A | 0.00 | 0.26 | 0.69 | 1.18 | 1.68 | 7.19 | 417 |
| | Region B | 0.01 | 0.61 | 1.12 | 1.81 | 2.75 | 9.60 | 153 |
| Maximum catchment slope | Globe | 0.03 | 2.91 | 9.20 | 12.73 | 19.71 | 66.48 | 599 |
| | Region A | 0.03 | 2.42 | 6.26 | 10.65 | 17.20 | 66.48 | 417 |
| | Region B | 0.45 | 7.88 | 14.89 | 17.46 | 24.58 | 43.90 | 153 |
| Mean topographic index | Globe | -0.25 | 1.80 | 2.35 | 2.34 | 2.90 | 5.10 | 593 |
| | Region A | -0.25 | 1.81 | 2.43 | 2.37 | 2.97 | 5.10 | 416 |
| | Region B | 0.76 | 1.70 | 2.20 | 2.19 | 2.63 | 3.95 | 152 |

Table B.3. Summary of the climate-based grouping of the examined catchments. The summarized information has been sourced from Do et al. (2018b). The term "non-classified" is here used to imply either that related information is not available for the catchment or that multiple climate types prevail in the catchment.

| Class | Globe | Region A | Region B |
|---|---|---|---|
| Arid | 32 | 29 | 0 |
| Equatorial | 7 | 0 | 0 |
| Non-classified | 8 | 1 | 6 |
| Polar | 29 | 0 | 29 |
| Snow | 371 | 276 | 95 |
| Warm Temperate | 152 | 111 | 23 |

Table B.4. Summary of the land-cover-based grouping of the examined catchments. The summarized information has been sourced from Do et al. (2018b). The term "non-classified" is here used as detailed in the caption of Table B.3.

| Class | Globe | Region A | Region B |
|---|---|---|---|
| Agriculture | 79 | 63 | 14 |
| Forest | 305 | 237 | 57 |
| Grassland | 23 | 15 | 4 |
| Non-classified | 143 | 63 | 74 |
| Settlement | 20 | 17 | 2 |
| Shrubland | 26 | 21 | 0 |
| Sparse vegetation | 1 | 0 | 1 |
| Wetland | 2 | 1 | 1 |

Table B.5. Summary of the lithology-based grouping of the examined catchments. The presented information has been sourced from Do et al. (2018b). The term "non-classified" is here used as detailed in the caption of Table B.3.

| Class | Globe | Region A | Region B |
|---|---|---|---|
| Acid plutonic rocks | 59 | 41 | 16 |
| Acid volcanic rocks | 8 | 4 | 4 |
| Basic plutonic rocks | 1 | 0 | 1 |
| Basic volcanic rocks | 15 | 14 | 0 |
| Carbonate sedimentary rocks | 57 | 46 | 11 |
| Intermediate plutonic rocks | 2 | 1 | 1 |
| Intermediate volcanic rocks | 2 | 1 | 0 |
| Metamorphics | 93 | 48 | 41 |
| Mixed sedimentary rocks | 42 | 22 | 18 |
| Non-classified | 103 | 54 | 41 |
| Pyroclastics | 1 | 0 | 0 |
| Siliciclastic sedimentary rocks | 167 | 154 | 7 |
| Unconsolidated sediments | 48 | 31 | 13 |
| Water Bodies | 1 | 1 | 0 |

Table B.6. Summary of the soil-based grouping of the examined catchments. The summarized information has been sourced from Do et al. (2018b). The term "non-classified" is here used as detailed in the caption of Table B.3.

| Class | Globe | Region A | Region B |
|---|---|---|---|
| Albeluvisols | 1 | 1 | 0 |
| Alisols | 8 | 8 | 0 |
| Andosols | 1 | 0 | 0 |
| Calcisols | 2 | 2 | 0 |
| Cambisols | 242 | 138 | 100 |
| Chernozems | 5 | 5 | 0 |
| Ferralsols | 11 | 0 | 0 |
| Kastanozems | 47 | 47 | 0 |
| Leptosols | 1 | 1 | 0 |
| Lixisols | 2 | 0 | 0 |
| Luvisols | 118 | 116 | 0 |
| Non-classified | 57 | 48 | 2 |
| Phaeozems | 7 | 7 | 0 |
| Podzols | 94 | 42 | 51 |
| Regosols | 2 | 2 | 0 |
| Vertisols | 1 | 0 | 0 |

Table B.7. Summary statistics for soil-profile characteristics of the examined catchments. The summarized information has been sourced from Do et al. (2018b).

| Variable | Region | Minimum | 1st quartile | Median | Mean | 3rd quartile | Maximum | Sample size |
|---|---|---|---|---|---|---|---|---|
| Bulk density (kg/m$^3$) | Globe | 184.2 | 395.1 | 506.3 | 526.0 | 641.9 | 1 174.5 | 599 |
| | Region A | 184.2 | 402.1 | 549.7 | 550.3 | 679.1 | 1 174.5 | 417 |
| | Region B | 190.6 | 379.7 | 451.0 | 449.5 | 521.4 | 692.8 | 153 |
| Clay content in soil profile | Globe | 1.50 | 5.35 | 7.86 | 8.18 | 10.40 | 21.82 | 599 |
| | Region A | 1.60 | 5.52 | 7.93 | 8.14 | 10.33 | 19.43 | 417 |
| | Region B | 1.50 | 4.66 | 7.05 | 7.27 | 9.50 | 15.12 | 153 |
| Sand content in soil profile | Globe | 1.89 | 13.63 | 18.93 | 19.29 | 24.00 | 44.83 | 596 |
| | Region A | 1.89 | 12.55 | 17.85 | 18.45 | 23.29 | 44.83 | 414 |
| | Region B | 7.39 | 17.03 | 20.67 | 20.70 | 24.50 | 36.95 | 153 |
| Silt content in soil profile | Globe | 2.63 | 13.27 | 17.74 | 17.77 | 21.83 | 41.75 | 599 |
| | Region A | 4.50 | 13.25 | 17.71 | 18.02 | 22.12 | 41.75 | 417 |
| | Region B | 8.11 | 15.67 | 18.77 | 18.62 | 21.85 | 31.45 | 153 |

Table B.8. Summary statistics for technical exploitation characteristics of the examined catchments. The number of dams is considered as a categorical variable in Appendix D; therefore, it is also summarized with Table B.9. The summarized information has been sourced from Do et al. (2018b).

| Variable | Region | Minimum | 1st quartile | Median | Mean | 3rd quartile | Maximum | Sample size |
|---|---|---|---|---|---|---|---|---|
| Number of dams | Globe | 0 | 0 | 0 | 3.21 | 2 | 119 | 599 |
| | Region A | 0 | 0 | 0 | 2.70 | 2 | 119 | 417 |
| | Region B | 0 | 0 | 0 | 3.90 | 1 | 102 | 153 |
| Total storage volumes (km³) | Globe | 0.00 | 0.00 | 0.00 | 1 938.10 | 263.50 | 310 774.00 | 599 |
| | Region A | 0.00 | 0.00 | 0.00 | 1 105.00 | 344.00 | 37 296.00 | 417 |
| | Region B | 0.00 | 0.00 | 0.00 | 374.50 | 157.00 | 9 872.00 | 153 |
| Mean catchment drainage density (km$^{-1}$) | Globe | 0.00 | 0.28 | 0.39 | 0.40 | 0.50 | 0.91 | 535 |
| | Region A | 0.00 | 0.26 | 0.39 | 0.40 | 0.50 | 0.91 | 417 |
| | Region B | 0.15 | 0.30 | 0.37 | 0.39 | 0.46 | 0.68 | 93 |
| Mean catchment irrigation area | Globe | 0.00 | 0.00 | 0.00 | 0.01 | 0.00 | 0.19 | 599 |
| | Region A | 0.00 | 0.00 | 0.00 | 0.01 | 0.01 | 0.19 | 417 |
| | Region B | 0.00 | 0.00 | 0.00 | 0.00 | 0.00 | 0.12 | 153 |

Table B.9. Summary of the grouping of the examined catchments according to the number of dams present within their boundaries. The summarized information has been sourced from Do et al. (2018b).

| Class | Globe | Region A | Region B |
|---|---|---|---|
| No dams | 349 | 234 | 98 |
| One dam | 85 | 62 | 20 |
| Two to four dams | 77 | 67 | 7 |
| Five to ten dams | 46 | 29 | 14 |
| More than ten dams | 42 | 25 | 14 |

**Appendix C    Comparison of forecasting methods in detail**

In this Appendix, we present in detail the results of our big data comparison of the five individual and 26 simple combination forecasting methods. In what follows, we pay particular attention to the assessment in terms of RMSE. Despite this focus, we also present and outline the results in terms of MAE, MAPE, MdAE and MdAPE. This multi-faced presentation is important, since different metrics could support different applications.

Starting from Figure 4, we first observe that, independently of the region and the metric, the benchmark has the worst average ranking across the examined river flow time series. It is, therefore, meaningful for someone to consider using the remaining 30 methods, which are more sophisticated and require longer records. In what follows, the benchmark will be also representing all weak methods that could be used as base methods under a simple forecast combination approach. For the entire dataset (composed by 599 time series; see the presentations under the label "Globe") and in terms of RMSE, the combiner of (1),(2),(4),(5) is the best-performing method with average ranking equal to 12.73, while the best-performing base method (i.e., the automatic ARFIMA method) has average ranking equal to 14.01. Independently of the region and the metric, most of the simple combination methods perform equally well or even better than each of their base methods.

To additionally assess how large the differences in predictive performance are between the methods, we present the average values of the scale-independent metrics (see Figure C.1) and the average relative improvements provided by all methods with respect to the benchmark in terms of the scale-dependent metrics (see Figure 5). The presentation is made conditional on the metric, the method and the region. By examining Figure C.1, we understand that all methods (with few exceptions) exhibit very close performance in terms of MAPE and MdAPE.

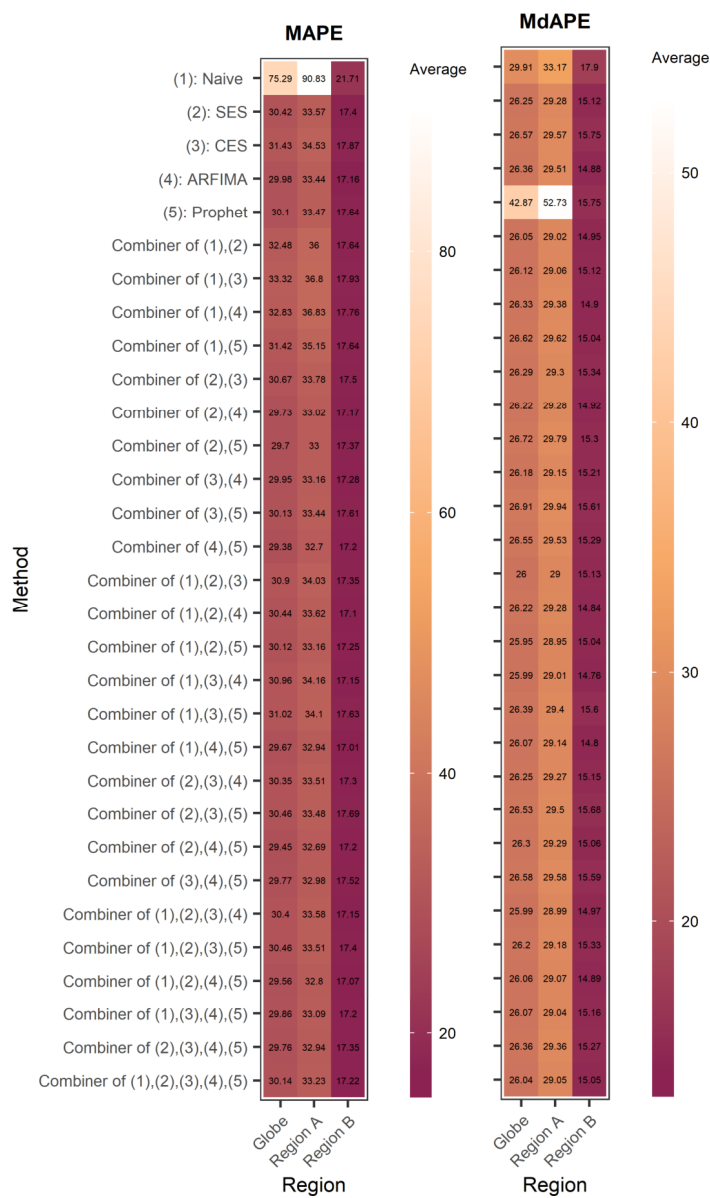

Figure C.1. Average MAPE and MdAPE values conditional on the metric, the method and the region. Each presented value concerning the entire dataset, Region A and Region B summarizes 599, 417 and 153 values, respectively.

The various forecasting methods differentiate more with each other in Figure 5. Perhaps the most important observations extracted from this latter figure concern the assessment in terms of RMSE (and MAE) and are the following: (i) All four combiners exploiting a well-performing method (i.e., only one of methods (2), (3), (4), (5) of the study) and a bad-performing one (i.e., method (1) of the study) improve over the performance of the bad-performing method from approximately 13% to approximately 17%; and (ii) all the remaining 22 simple combination methods are at least as good (roughly), and usually better, than each of the base methods.

The combiner of (1),(4),(5) is the best-performing method in terms of RMSE and for the entire dataset. This method combines the results provided by the benchmark, ARFIMA and Prophet. It improves over the performance of the benchmark on average by 18.9%, while the respective average relative improvement provided by the best-performing base method is 18.0%. This could be important in technical applications that rely on accuracy in the long run, given that the remaining three well-performing base methods provide improvements that are equal to 16.9%, 15.7% and 14.6%.

For Region A, the average respective improvement in terms of RMSE offered by the best-performing simple combination method is 19.2%. This method is again the combiner of (1),(4),(5), while the best-performing base method improves over the performance of the benchmark by 18.1%. Finally, for Region B the best-performing simple combination methods are the combiner of (1),(4),(5) (again) and the combiner of (1),(2),(4),(5). These methods respectively improve over the performance of the benchmark on average by 20.4% and 20.3% in terms of RMSE, while the respective improvements offered by the best- and worst-performing base methods (the latter identified by excluding the benchmark) are 20.0% and 16.9%, respectively. It is relevant to note, at this point, that both the aforementioned simple combination methods exploit the forecasts by Naïve, ARFIMA and Prophet.

To further increase our understanding on how simple combination methods can be used to reduce risk in hydrological forecasting and other geoscientific and environmental contexts, we qualitatively present the normalized densities of the rankings conditional on the method and the metric (see Figures C.2 and C.3). By collectively examining Figures 6, 7, C.2 and C.3, we first observe that, independently of the metric, the benchmark is ranked at the last few positions for many of the 599 examined river flow time series. Especially in terms of MAE, MAPE and RMSE, rankings that are equal to 31 are by far the most frequent ones. Second, we observe that the performance of the remaining base methods (i.e., SES, CES, ARFIMA and Prophet) is largely varying from case to case with the very good and very bad performances (roughly represented by rankings 1–10 and 21–31, respectively) being more frequent than medium performances (roughly represented by rankings 11–20).

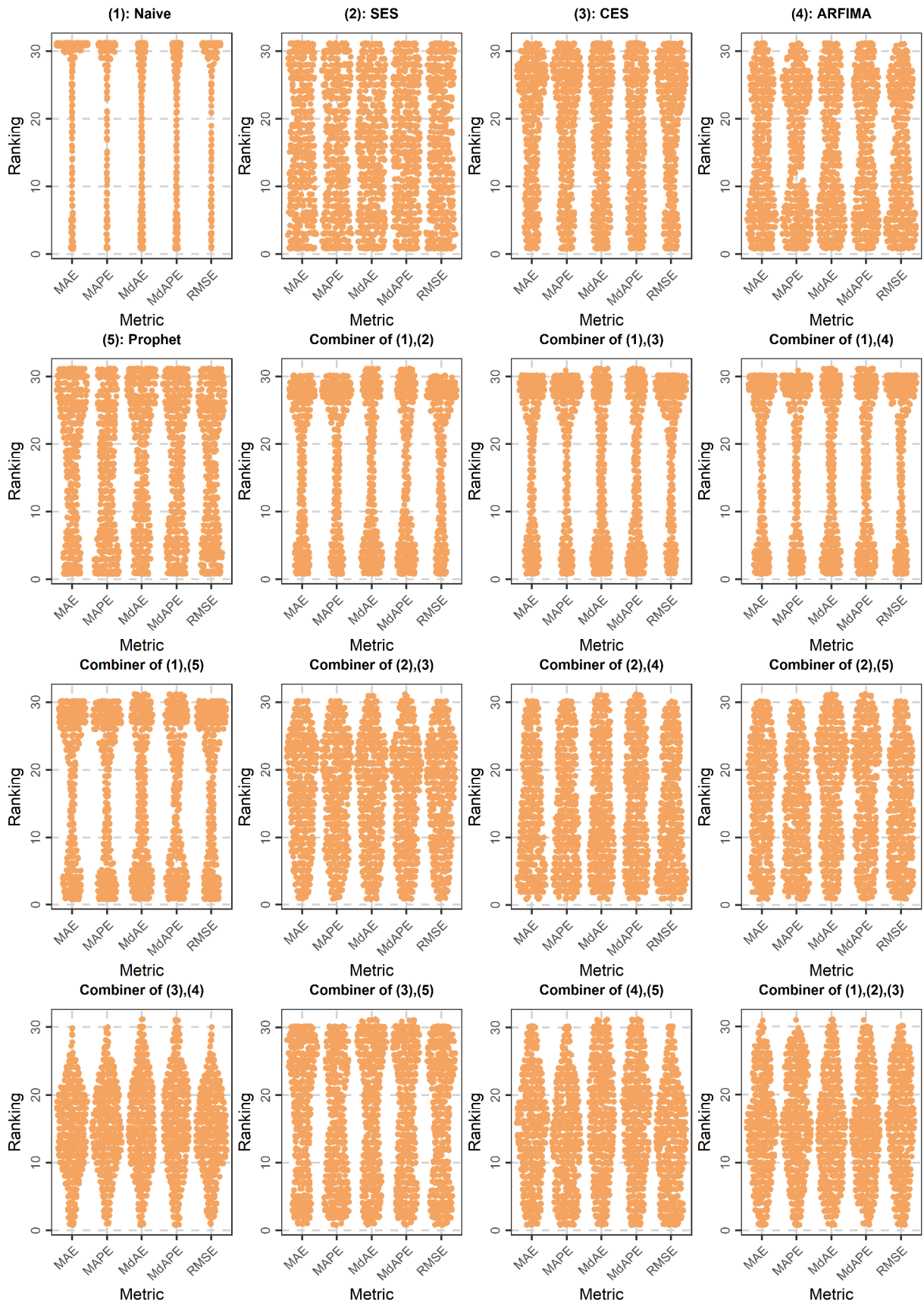

Figure C.2. Sinaplots of the rankings of the methods in all conducted tests conditional on the metric (part 1). The lower the ranking the better the performance.

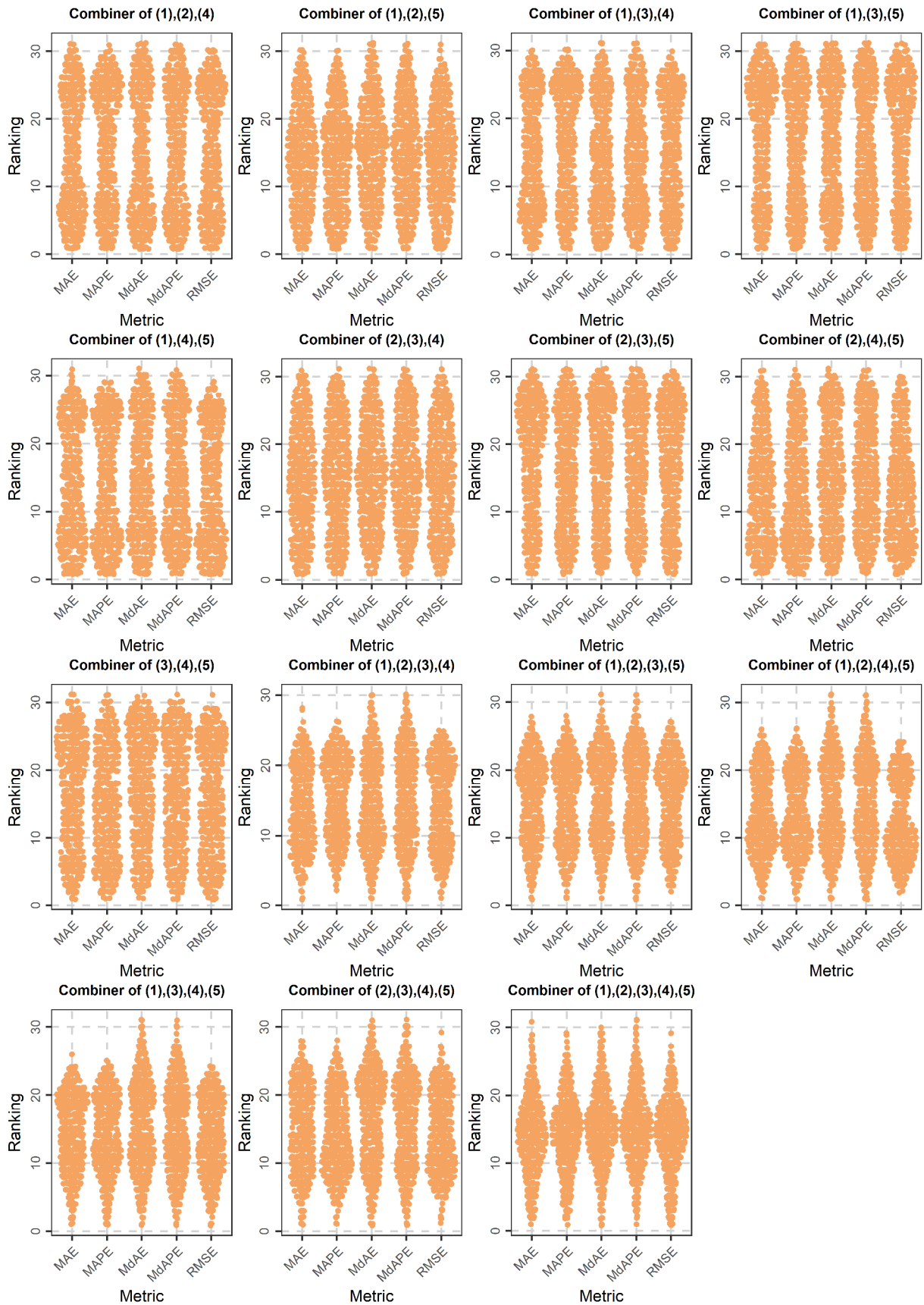

Figure C.3. Sinaplots of the rankings of the methods in all conducted tests conditional on the metric (part 2). Methods (1), (2), (3), (4) and (5) are defined in Figure C.2. The lower the ranking the better the performance.

Here again it is useful to roughly classify the 26 variants of the simple combination methodology of the study into two main groups. The first group includes the variants that combine the benchmark with one of the remaining base methods. In terms of MAE, MAPE and RMSE, these variants provide improvements with respect to the benchmark and deteriorations with respect to the other base method. In fact, while a significant portion of their forecasts is ranked in the ten first positions, most of their forecasts are ranked in the last ten positions. Fewer forecasts are of medium quality. On the other hand, in terms of MdAE and MdAPE most of their forecasts are ranked in the first five and the last five positions. A smaller portion is ranked from $6^{th}$ to $10^{th}$ and from $21^{st}$ to $26^{th}$ (roughly). Medium performances are also less frequent for these metrics.

The second group includes all simple combination methods that combine either (i) only base methods other than the benchmark or (ii) the benchmark with at least two other base methods. Different patterns are created by the various methods falling into this group. Nonetheless, a common characteristic of most of these methods is that they tend to produce forecasts that are ranked far from some few best positions but also far from some few worst positions (therefore, for them orange mostly dominates white and magenta in Figure 7). A representative example is the combiner of (3),(4). This method mostly delivers forecasts that could be characterized as good to medium in terms of rankings. Other good examples are all combiners combining any four or five methods, which manage to mostly eliminate forecasts that are ranked in the last five positions with a concomitant profit in average-case performance.

Lastly, our big data predictive performance results could be examined as a collection of 599 case studies (see e.g., the rankings of the methods conditional on the metric in 12 selected case studies, as presented in Figures C.4 and C.5). The entire collection can be found online at: https://doi.org/10.17632/z2mdnnxghg.1.

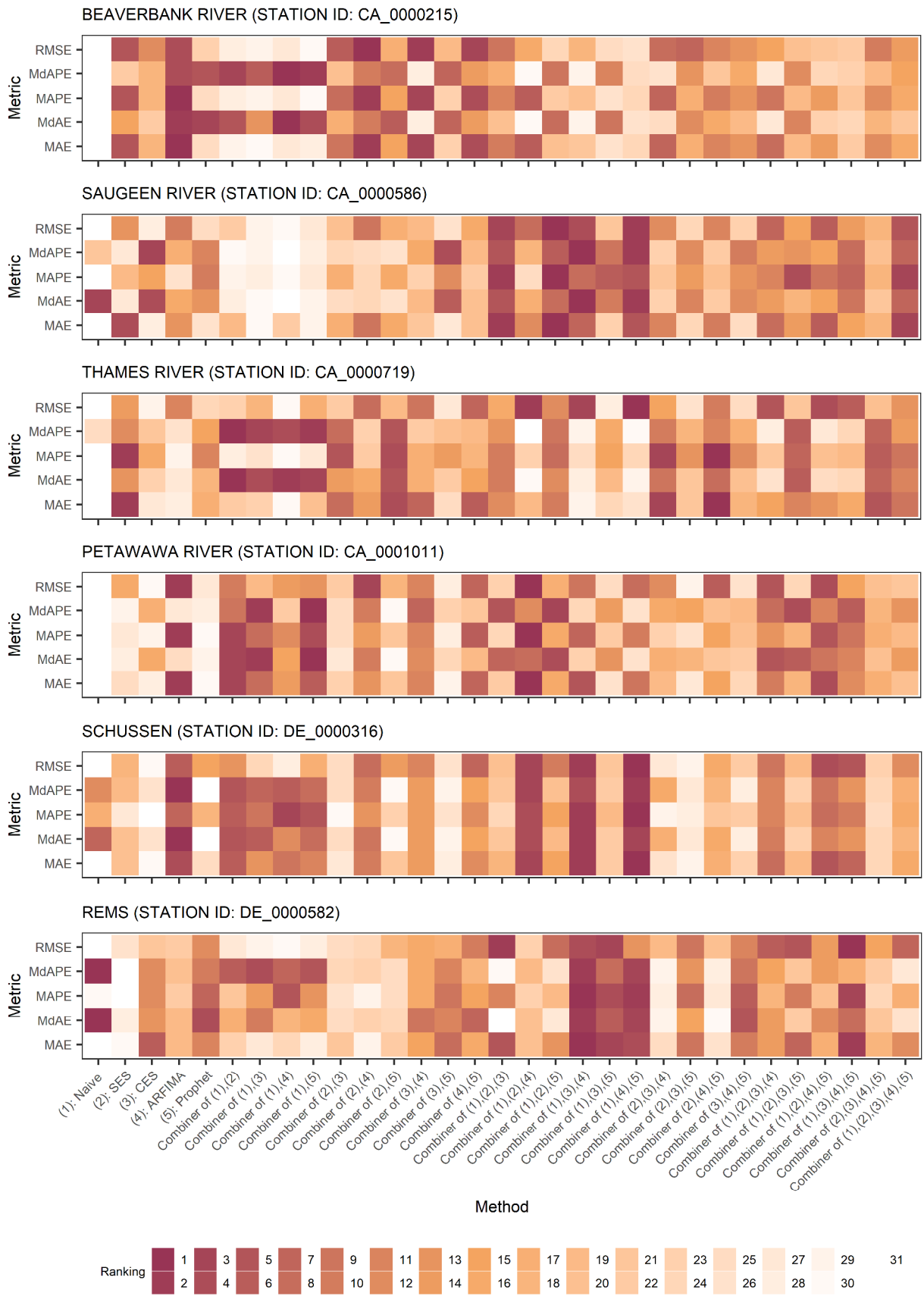

Figure C.4. Rankings of the methods in selected case studies conditional on the metric (part 1).

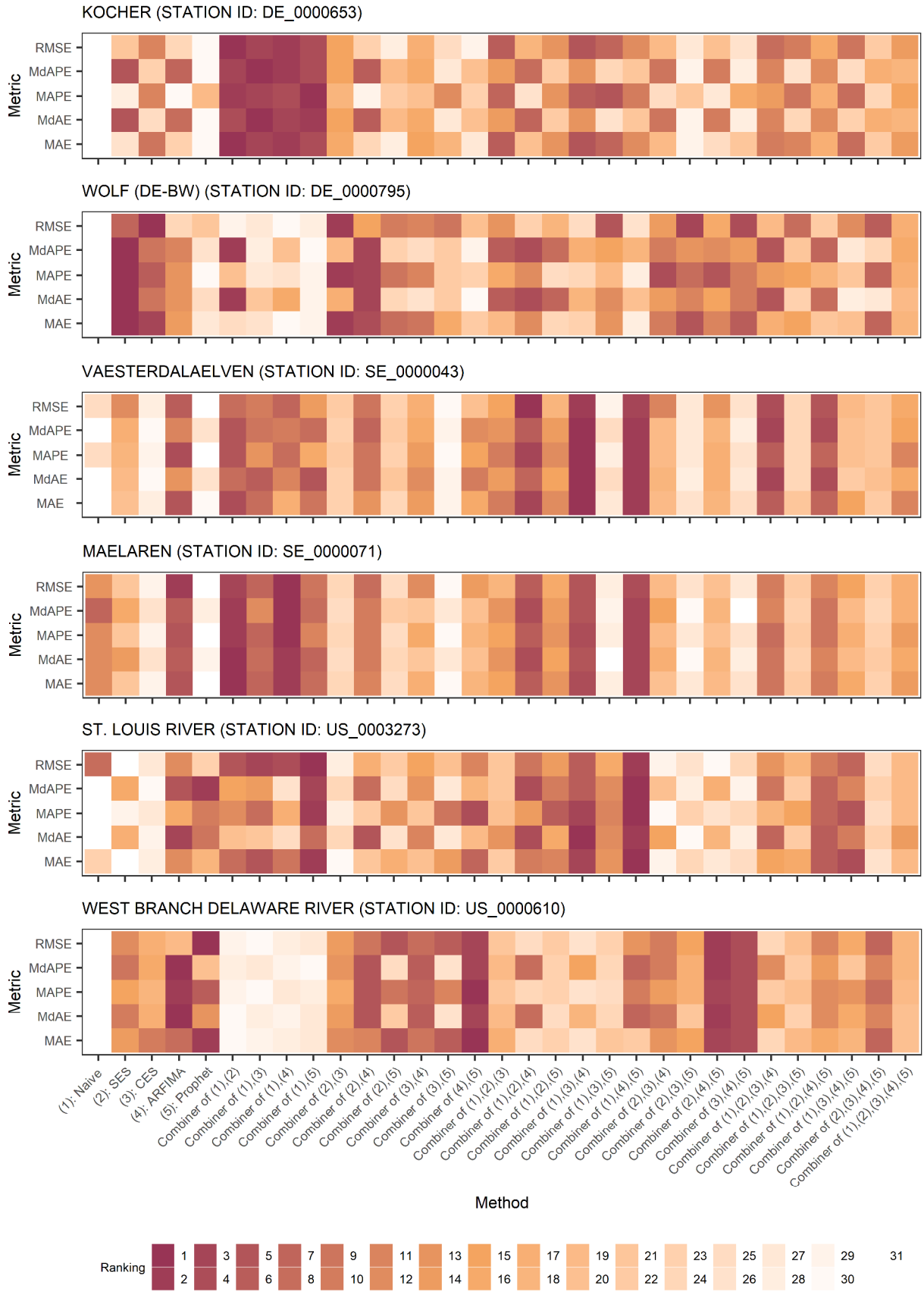

Figure C.5. Rankings of the methods in selected case studies conditional on the metric (part 2).

**Appendix D    Forecasting performance versus catchment location and attributes**

To further investigate the possibility of case-dependent exploitations of hydrological time series forecasting methods, in this Appendix we conduct investigations analogous to those presented in Section 3.3, but this time aiming to identify possible relationships between the relative improvements provided by all methods with respect to Naïve and selected catchment attributes (see Appendix B). The outcomes of these investigations are presented with Table D.1 and Figures D.1–D.8. We did not find any links between forecasting performance, and catchment location and attributes, probably due to the forecast time scale. Nonetheless, Figures D.1–D.8 collectively contribute (together with Figures C.4 and C.5) to making the presentation of the results of the forecasting tests more focused to the case itself.

Table D.1. Summary of the results of the linear regression analyses between the relative improvements in terms of RMSE provided by each method with respect to the benchmark for each of the 599 catchments (dependent variable) and each of the listed explanatory variables.

| Explanatory variable | Information | Slope | Pearson's $r$ | Illustrations |
|---|---|---|---|---|
| Station longitude (°) | Figure 1 | −0.01 | 0.04 | Figure D.1 |
| Station latitude (°) | | 0.19 | 0.16 | |
| Station altitude (m) | Table B.2 | −0.01 | 0.15 | – |
| Catchment area (km$^2$) | | 0.00 | 0.06 | |
| Estimated catchment area (km$^2$) | | 0.00 | 0.06 | |
| Mean catchment elevation (m) | | −0.01 | 0.13 | |
| Maximum catchment elevation (m) | | 0.00 | 0.14 | |
| Mean catchment slope | | 0.02 | 0.00 | |
| Maximum catchment slope | | −0.13 | 0.08 | |
| Mean topographic index | | −1.44 | 0.07 | |
| Bulk density (kg/m$^3$) | Table B.7 | −0.01 | 0.09 | Figure D.2 |
| Clay content in soil profile | | −0.85 | 0.16 | |
| Sand content in soil profile | | −0.03 | 0.01 | |
| Silt content in soil profile | | 0.15 | 0.05 | |
| Total storage volumes (km$^3$) | Table B.8 | 0.00 | 0.03 | – |
| Mean catchment drainage density (km$^{-1}$) | | −0.24 | 0.00 | |
| Mean catchment irrigation area | | −102.91 | 0.12 | |
46

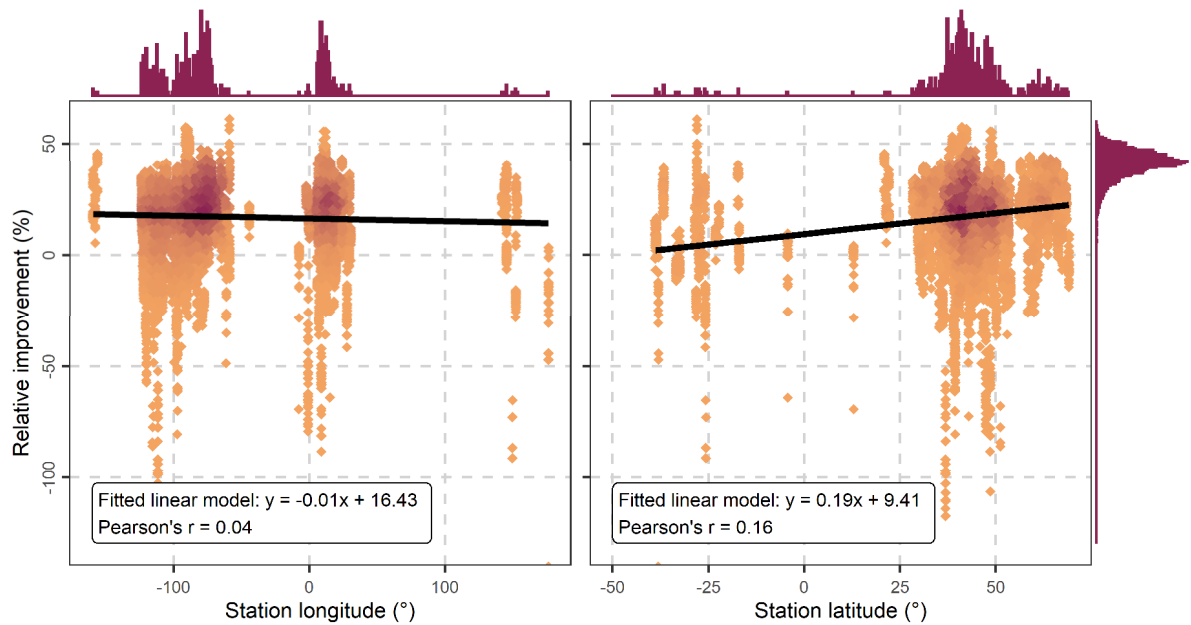

Figure D.1. Aggregated relative improvements in terms of RMSE provided by each method with respect to the benchmark for each of the 599 catchments (30 × 599 = 17 970 values) in comparison to the station longitude (left panel) and the station latitude (right panel). Station longitude and latitude information is presented in Figure 1. Orange and magenta data points respectively denote low and high density, while the black lines denote the linear models fitted to these data points. The vertical axes have been truncated at −130.

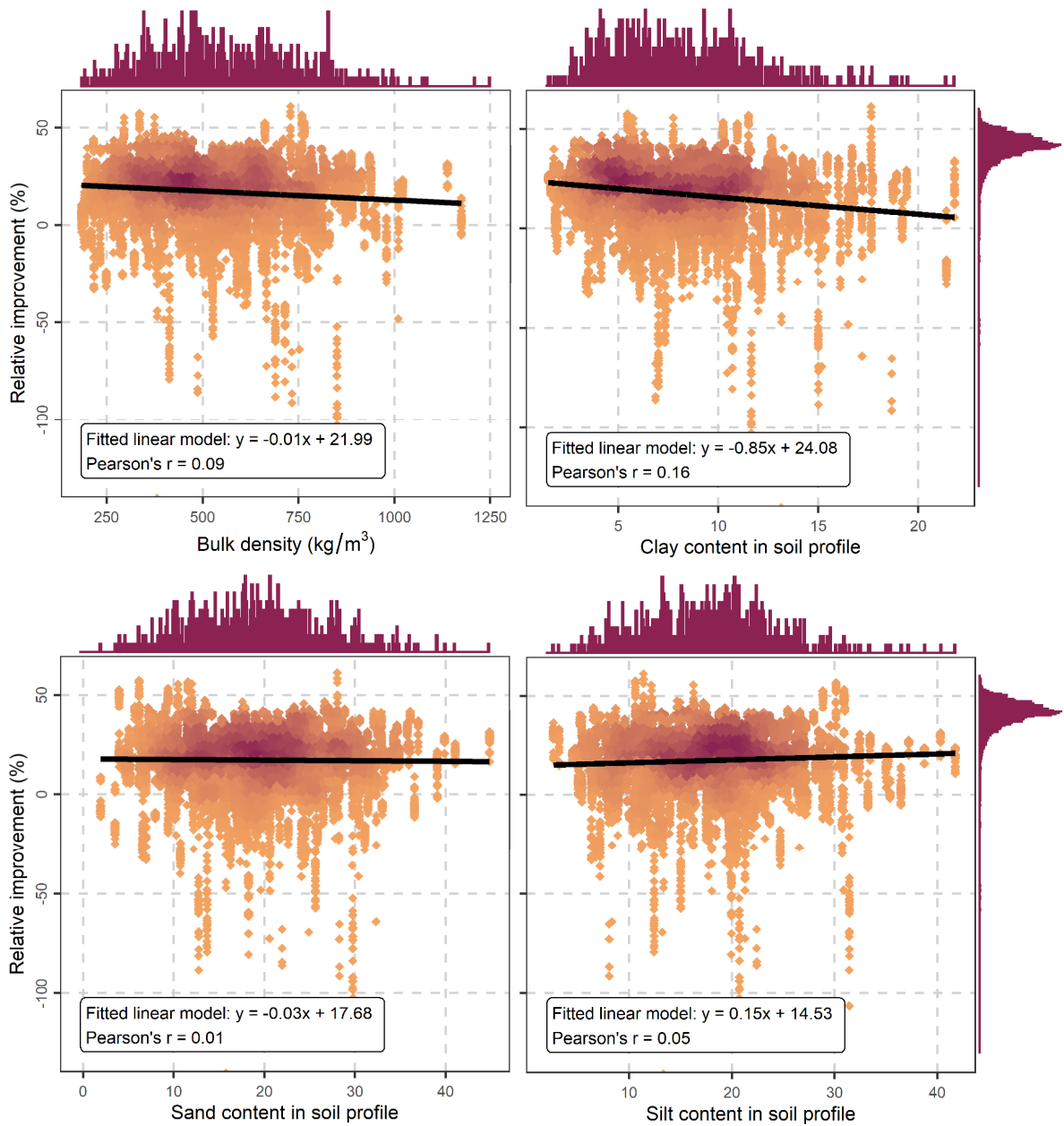

Figure D.2. Aggregated relative improvements in terms of RMSE provided by each method with respect to the benchmark for each of the 599 catchments (30 × 599 = 17 970 values) in comparison to the bulk density (upper left panel), the clay content in soil profile (upper right panel), the sand content in soil profile (lower left panel) and the silt content in soil profile (lower right panel). Bulk density, clay content, sand content and silt content information is summarized in Table B.7. Orange and magenta data points respectively denote low and high density, while the black lines denote the linear models fitted to these data points. The vertical axes have been truncated at −130.

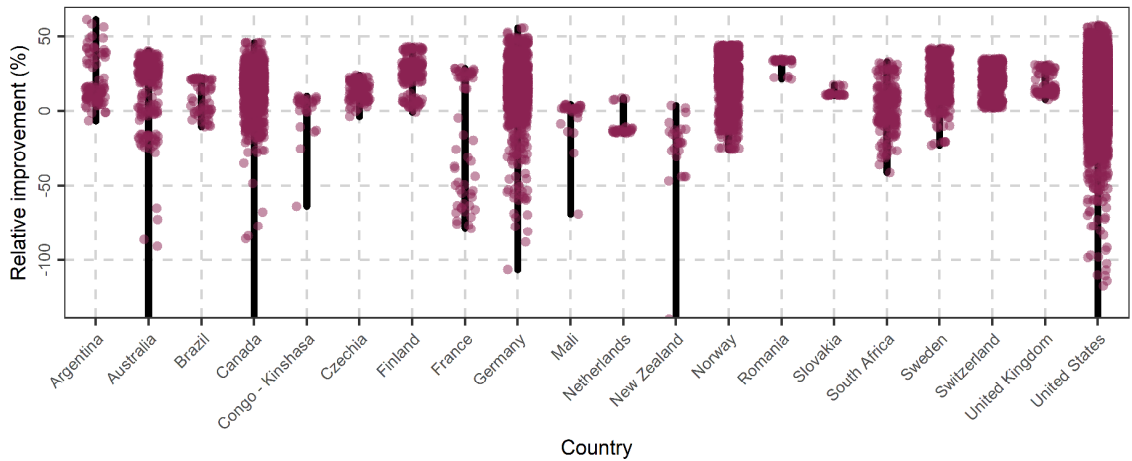

Figure D.3. Aggregated relative improvements in terms of RMSE provided by each method with respect to the benchmark for each of the 599 catchments (30 × 599 = 17 970 values) conditional on the country. The number of catchments located to each country is presented in Table B.1. The relative improvements are presented as jitter points. Violin areas (denoted with black colour) are scaled proportionally to the number of relative improvements. The vertical axis has been truncated at −130.

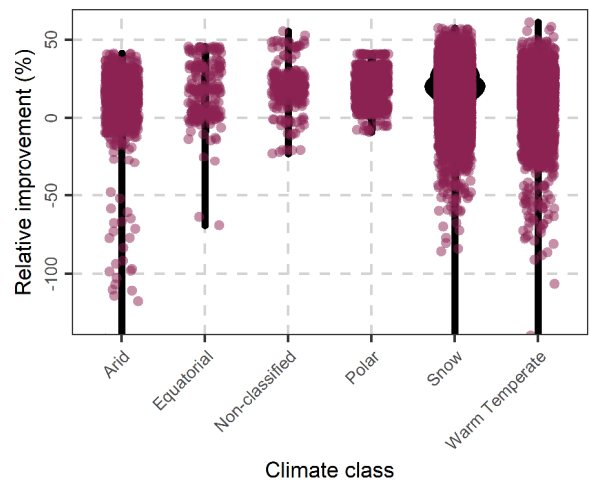

Figure D.4. Aggregated relative improvements in terms of RMSE provided by each method with respect to the benchmark for each of the 599 catchments (30 × 599 = 17 970 values) conditional on the climate class. The number of catchments assigned to each class is presented in Table B.3. The relative improvements are presented as jitter points. Violin areas (denoted with black colour) are scaled proportionally to the number of relative improvements. The vertical axis has been truncated at −130.

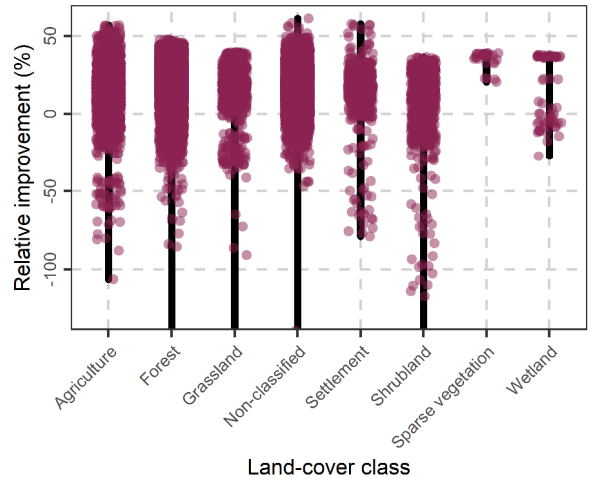

Figure D.5. Aggregated relative improvements in terms of RMSE provided by each method with respect to the benchmark for each of the 599 catchments (30 × 599 = 17 970 values) conditional on the land-cover class. The number of catchments assigned to each class is presented in Table B.4. The relative improvements are presented as jitter points. Violin areas (denoted with black colour) are scaled proportionally to the number of relative improvements. The vertical axis has been truncated at −130.

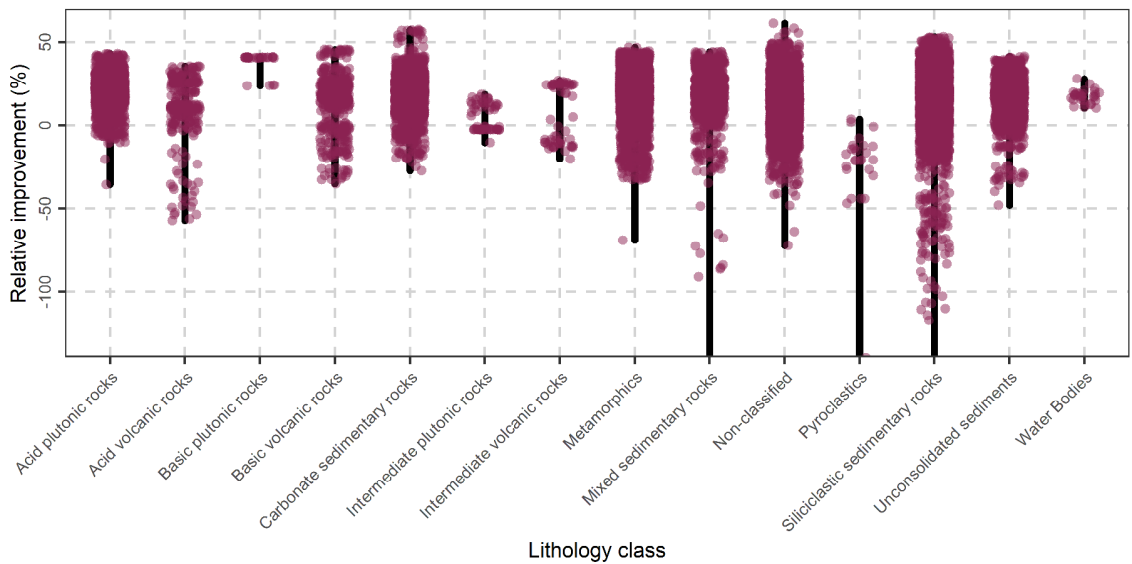

Figure D.6. Aggregated relative improvements in terms of RMSE provided by each method with respect to the benchmark for each of the 599 catchments (30 × 599 = 17 970 values) conditional on the lithology class. The number of catchments assigned to each class is presented in Table B.5. The relative improvements are presented as jitter points. Violin areas (denoted with black colour) are scaled proportionally to the number of relative improvements. The vertical axis has been truncated at −130.

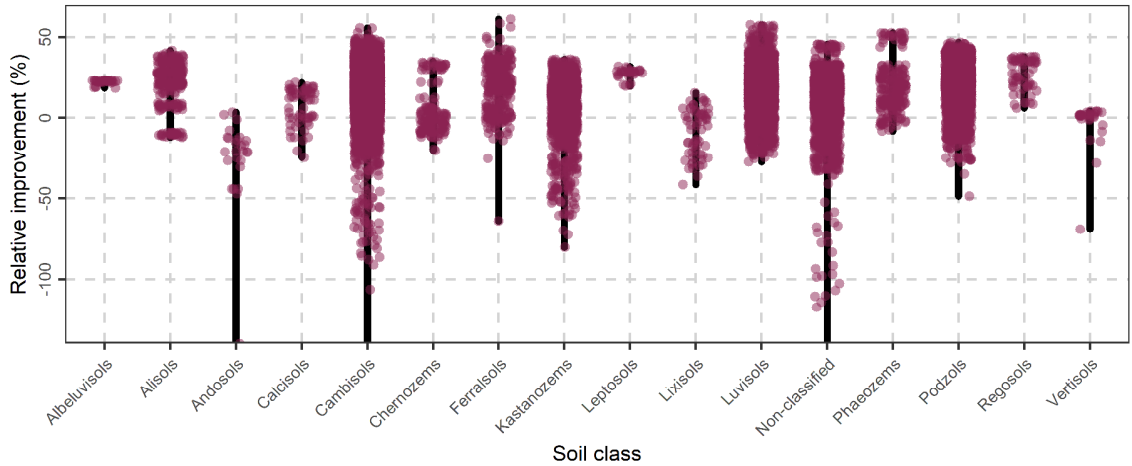

Figure D.7. Aggregated relative improvements in terms of RMSE provided by each method with respect to the benchmark for each of the 599 catchments (30 × 599 = 17 970 values) conditional on the soil class. The number of catchments assigned to each class is presented in Table B.6. The relative improvements are presented as jitter points. Violin areas (denoted with black colour) are scaled proportionally to the number of relative improvements. The vertical axis has been truncated at −130.

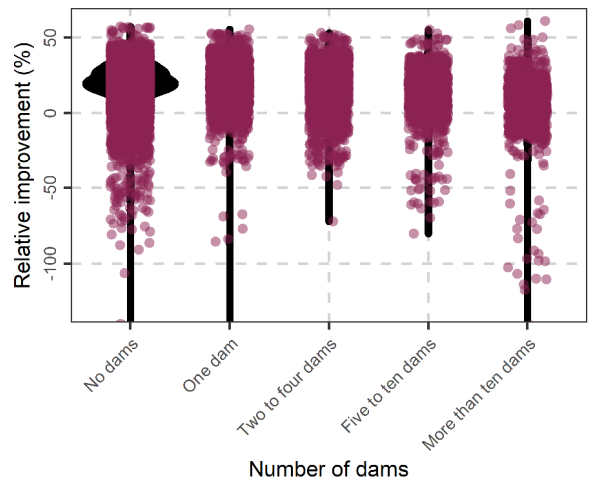

Figure D.8. Aggregated relative improvements in terms of RMSE provided by each method with respect to the benchmark for each of the 599 catchments (30 × 599 = 17 970 values) conditional on the number of dams present in the catchments' boundaries. The number of catchments assigned to each class is presented in Table B.9. The relative improvements are presented as jitter points. Violin areas (denoted with black colour) are scaled proportionally to the number of relative improvements. The vertical axis has been truncated at −130.

**Declarations of interest:** We declare no conflict of interest.